\documentclass[sigconf]{acmart}
\AtBeginDocument{%
  }

\usepackage{rotating}
\usepackage{array}
\usepackage{colortbl}
\usepackage{booktabs}
\usepackage{tablefootnote}
\usepackage[caption=false,font=footnotesize]{subfig}
\usepackage{wrapfig}

\usepackage[flushleft]{threeparttable}

\usepackage{comment}
\usepackage{amsfonts}
\usepackage{algorithmic}
\usepackage{graphicx}
\usepackage{textcomp}
\usepackage{booktabs}
\usepackage{tabularx}
\usepackage{booktabs}
\usepackage{multirow} % for table
\usepackage{graphicx}
\usepackage{float}
\usepackage{lipsum}
\usepackage{msc}
\usepackage{xcolor}
\usepackage{tcolorbox} 
\usepackage{soul} % use for highlight hl{..}
\usepackage{enumitem}
\usepackage{pifont}
\usepackage{graphicx}

\newtcolorbox{roundedtextbox}[1][]{%
  colback=black!5!white,
  colframe=black!40!white,
  fonttitle=\bfseries,
  arc=10pt,  % Controls the roundness of the corners
  boxrule=1pt,
  fontupper=\small,
  #1
}

\usepackage{circledsteps}
\usepackage{tikz}
\usepackage{xcolor}
\pgfkeys{/csteps/inner ysep=2pt}
\pgfkeys{/csteps/inner xsep=5pt}
\newcommand*\circled[1]{\tikz[baseline=(char.base)]{
            \node[shape=circle,fill,inner sep=0pt] (char) {\textcolor{white}{#1}};}}

\newcommand{\implementationcount}{19}
\newcommand{\modelcount}{186} 
\newcommand{\referenceModelCount}{11}

\newcommand{\bugcount}{55}
\newcommand{\protocolViolation}{44}
\newcommand{\softwareBugs}{8}
\newcommand{\logicBugs}{3}
\newcommand{\totalsymb}{30}

\newcommand{\basic}{\textbf{Basic}}

\newcommand{\clientAuthentication}{\textbf{ClientAuth}}
\newcommand{\retryClientAuthentication}{\textbf{RetryClientAuth}}
\newcommand{\psk}{\textbf{PSK}}
\newcommand{\supp}{\textit{Appendix}}

\newcommand{\learner}{{\textsf{Learner}}}
\newcommand{\mapper}{{\textsf{Test Harness}}}
\newcommand{\sut}{{\textsf{QUT}}}
\newcommand{\opt}{{\textsf{Optimizer}}}
\newcommand{\differentialComparator}{{\textsf{Differential Analyzer}}}
\newcommand{\logger}{{\textsf{Crash Logger}}}

\newcommand{\add}[1]{\textcolor{blue}{#1}}
\usepackage[normalem]{ulem}

\definecolor{BackgroundBlue}{RGB}{218, 232, 252}
\definecolor{BackgroundYellow}{RGB}{255, 242, 204}
\definecolor{BackgroundOrange}{RGB}{255, 230, 204}
\definecolor{BackgroundRed}{RGB}{248, 206, 204}
\definecolor{BackgroundGreen}{RGB}{213, 232, 212}
\definecolor{BackgroundPurple}{RGB}{225, 213, 231}

\usepackage[T1]{fontenc}
\newcommand{\pmsg}[1]{{\fontfamily{lmss}\selectfont{#1}}}
\newcommand{\pval}[1]{{\fontfamily{QTHelvetCnd}\selectfont{#1}}}

\newcommand{\name}{\textsc{QUICtester}}

\usepackage{xcolor}
\usepackage{tcolorbox}
\usepackage[framemethod=TikZ]{mdframed}

\begin{document}

\title[\name{}: An Automated Noncompliance Checker for QUIC]{An Automated Blackbox Noncompliance Checker for QUIC Server Implementations}

\author{Kian Kai Ang}
 \affiliation{
\institution{The University of Adelaide}
\country{Australia}
}
\email{kiankai.ang@adelaide.edu.au}

\author{Guy Farrelly}
 \affiliation{
\institution{The University of Adelaide}
\country{Australia}
}
\email{guy.farrelly@adelaide.edu.au}

\author{Cheryl Pope}
 \affiliation{
\institution{The University of Adelaide}
\country{Australia}
}
\email{cheryl.pope@adelaide.edu.au}

\author{Damith C. Ranasinghe}
 \affiliation{
\institution{The University of Adelaide}
\country{Australia}
}
\email{damith.ranasinghe@adelaide.edu.au}

\begin{abstract}
We develop \name{}, an automated approach for uncovering non-compliant behaviors in the ratified QUIC protocol implementations (RFC 9000/9001). \name{} leverages active automata learning to abstract the behavior of a QUIC implementation into a finite state machine (FSM) representation. Unlike prior noncompliance checking methods, to help uncover state dependencies on event timing, \name{} introduces the idea of state learning with event timing variations, adopting both valid and invalid input configurations, and combinations of security and transport layer parameters during learning. We use pairwise differential analysis of learned behaviour models of tested QUIC implementations to identify non-compliance instances as behaviour deviations in a property-agnostic way. This exploits the existence of the many different QUIC implementations, removing the need for validated, formal models. The diverse implementations act as cross-checking test oracles to discover non-compliance. We used \name{} to analyze \modelcount{} learned models from \implementationcount{} QUIC implementations under the \textit{five} security settings and discovered \bugcount{} implementation errors. Significantly, the tool uncovered a QUIC specification ambiguity resulting in an easily exploitable DoS vulnerability, led to 5 CVE assignments from developers, and two bug bounties thus far.

\vspace{2mm}
\noindent\textbf{Code \& PoCs}: \color{blue}\textsf{\url{https://github.com/QUICTester}}.\color{black}
\end{abstract}




\maketitle

\section{Introduction}
Ratified in May 2021, QUIC is a performance-optimized, secure, reliable transport protocol for the Internet and a core part of the HTTP/3 protocol. QUIC is a ground up re-design aiming to reduce latency and connection overhead associated with the use of TLS (Transport Layer Security)~\cite{RN49} over TCP for secure transport~\cite{RN43} and achieve \textit{inherently} secure communication channels---ensuring message confidentiality, integrity, and availability for Internet applications.  According to~\cite{QUICstats}, as of November 2023, 27\% of all websites use HTTP/3 employing QUIC for the transport protocol---including Google, Meta, Amazon and all major browsers---with use cases also extending to the Domain Name System (DNS)~\cite{9250}. Further, with the significant growth in Internet of Things applications, projected to be more than 29 billion by 2027~\cite{state_of_iot}, we can expect QUIC to dominate secure, reliable data transfer over the  future Internet~\cite{iotQuicProve, kumar2019implementation, hou2022qfaas}.

Despite significant efforts to investigate secure protocols---such as TLS~\cite{somorovsky2016systematic,tlsFuzzer,6956560,7163046}, Datagram Transport Layer Security (DTLS)~\cite{dtlsFuzzer, fiterau2023automata}, OpenVPN~\cite{RN21} and the 802.11 4-Way Handshake~\cite{RN22}---there is a gap in reliable \textit{tools} to scrutinize specification conformance of QUIC implementations and consequential vulnerabilities. Our motivation is to address this gap. 

Prior to ratifying the QUIC specification~\cite{9000,9001}, early efforts made advances to develop methods and tools to validate QUIC implementations~\cite{RN15, crochet2021verifying, kakhki2017taking, mcmillan2019formal}. But, the tools are specific to Google-QUIC, are no longer actively maintained, not open source or are limited in their scope and suitability for evaluating the ratified QUIC protocol as we discuss in Section~\ref{sec:discussion}. An effective and noncompliance testing method is important for verifying the security promised by QUIC is delivered by implementations. 

In the absence of an open-source testing method for the QUIC specification~\cite{9000,9001} to identify non-compliances and potential security vulnerabilities, \textit{our work presents the first open-source, comprehensive design and implementation of a framework for analyzing implementations of the ratified QUIC standard}.

\vspace{2mm}
\noindent\textbf{Our Work.~}Our goal is to provide a tool that automates the task of uncovering: i)~non-conforming protocol behaviors; and ii)~security vulnerabilities exploitable by crafting specific message sequences. To this end, we design and build \name{}---a comprehensive, automated, black-box tester for uncovering non-compliant behaviors and security vulnerabilities in QUIC implementations. Notably, \name{}  has the desirable property of being agnostic to the QUIC implementation and run-time environment, such as the programming language, operating system, and CPU instruction set architecture. We address a number of key challenges in realizing an effective \textit{QUIC-specific} noncompliance checker. 

\vspace{2mm}
\noindent\textit{(1)~Automatically Learning a Behavior Model Under Valid and Invalid Inputs.~}Analyzing non-conforming behavior of a protocol implementation requires construction of a model of the underlying implementation and comparison with a formal model of a protocol by a domain expert. The process requires a significant manual effort and scaling the effort to a large number of implementations is often impractical. In addition to potential human error, the process is made more difficult given specification ambiguities, under-specification, the length of specifications (the QUIC specification spans 80,000 words incorporating five different security levels, as well as several security and transport parameter options). 

Our approach uses active automata learning to overcome the lengthy task of eliciting the behavior abstractions of a QUIC implementation to a finite state machine (FSM) representation. Active automata learning aims to infer a system's state space and possible transitions between those states by sending a series of input sequences to the system and observing the corresponding outputs~\cite{modelLearning, dynamicAutomataLearning}. In the process, all input sequences and corresponding output sequences explored in a model learning phase are inferred into an FSM description of the target system. Although past efforts used active automata learning to build testing tools for network protocols~\cite{dtlsFuzzer,tlsFuzzer,fiteruau2016combining,RN17,mqttFuzzer,RN51,RN21,RN22,hussain2021noncompliance}, an active learner for the QUIC specification does not yet exist. We design such a \learner{}.

To allow automata learning algorithms to generate tests---a series of input sequences, \textit{we construct the first comprehensive QUIC-specific \learner{} for all secure handshake configurations through careful examination of the QUIC RFC~9000/9001. The result is a highly expressive learner capable of generating both valid and \textit{invalid} packet configurations as well as various security and transport parameter combinations in the learning phase.}

\vspace{2mm}
\noindent\textit{(2)~Learning Time-Dependent Behavior Models.~}Further, we recognize, network protocols like QUIC, are typically time sensitive due to timeout behaviors.  Therefore, we can expect timing variations, such as intervals between packet transmissions, to impact protocol behavior and potentially expose hidden states and weaknesses in protocol implementations that may be exploitable. In contrast to prior methods to analyze protocols, \textit{we introduce the idea of exploring temporal dependencies on protocol states. We parameterize the symbols with time to allow the learner to self-select timing variations}.

\vspace{2mm}
\noindent\textit{(3)~Test Harness.~}Because the learning algorithms are protocol agnostic and only operate on the symbols and parameters, a protocol-specific test harness is needed to translate parameterized symbols from a learner into protocol messages and vice versa.  This harness manages the interactions between a model \learner{} and the target QUIC implementation under test.  

The complexity of the QUIC protocol---supporting 5 security levels, with the combination of transport parameters and cryptographic negotiations to manage various types of packets and frames---makes building a test harness a significant challenge. \textit{We built a comprehensive QUIC protocol test harness to enable testing all, five secure handshake configurations in QUIC implementations}.
 
\vspace{2mm}
\noindent\textit{(4)~Automating Analysis.~}A significant challenge is to identify non-compliant behavior. To reduce the cumbersome, error-prone, manual effort in the analysis phase, we combine two strategies to achieve an effective automated analysis method. \textit{First}, we propose a set of \textit{optimizations} to eliminate redundant information from learned models while preserving the captured behavior. \textit{Second}, we construct a differential analysis method---pair-wise testing---to automatically identify non-conforming behaviors. Differential analysis simplifies the task of identifying non-compliant behavior to the task of identifying \textit{deviating behaviors} based on comparing models against each other. Through the results of differential testing and our manual analysis and validation, we are able to contribute a curated library of reference FSM models for the QUIC specification for each security configuration. The models represent behavior that conforms to the specification. \textit{These models serve developers to employ \name{} to effectively and efficiently identify non-conformance behaviors of a target implementation} (see Fig.~\ref{fig:QUIC-TeSter-use-case}).

\vspace{2mm}
\noindent\textbf{Scope.~}We focus our tests on the more impactful server-side implementations of QUIC as server failures affect multiple active connections; for example, Denial of Service attacks. Notably, the same protocol library is used by clients and servers. Further, our focus is on security. Hence, we test the handshake component \textit{crucial} for establishing secure, multi-stream, connections with QUIC.

\vspace{2mm}
\noindent\textbf{Contributions.~}In this work, in summary:
\begin{itemize}[itemsep=2pt,parsep=1pt,topsep=3pt,labelindent=0pt,leftmargin=5mm]
    \item We propose \name{}, a blackbox testing framework for QUIC to \textit{automatically} identify specification deviation (non-compliance checking), uncover logical flaws (functional bugs) and security vulnerabilities without needing a formal reference model description. 
    
    \item We introduce the idea of learning a model under \textit{event timing variations}. We craft a conceptually simple, yet protocol-agnostic means to achieve it.
    
    \item We design and implement a comprehensive QUIC-specific \mapper{} adhering to RFC 9000/9001 to support the complex \learner{} sequence combinations (valid, invalid, time dependent message with varying protocol parameters, both transport and security) and all of the security configurations to test QUIC implementations---see Table~\ref{tab:quicImplementations}.
            
    \item To alleviate the manual analysis burden, we automate analysis of learned models. We manually validate our method with \modelcount{} learned models and curate conforming models for various security configurations to serve as reference models. The reference models with \name{} can support practitioners' use of our tool to efficiently test and analyze QUIC implementation targets (see Section~\ref{sec:discussion} and Fig.~\ref{fig:QUIC-TeSter-use-case}).
    
    \item We open-source \name{}, at \color{blue}\textsf{\url{https://github.com/QUICTester}}.\color{black}
\end{itemize}

\vspace{2mm}
\noindent\textbf{Findings.~}We used \name{} with publicly available QUIC server releases to assess its effectiveness and help improve security and interoperability of QUIC implementations. 
\begin{itemize}
[itemsep=2pt,parsep=1pt,topsep=4pt,labelindent=0pt,leftmargin=5mm]
    \item We test \implementationcount{} QUIC implementations summarized in Table~\ref{tab:quicImplementations} under the \textit{five} different \textit{security} configurations from a range of providers.  We analyzed \modelcount{} learned models to uncover \bugcount{} faults. Our work has been validated with a \textit{bug bounty award} and 5 CVE assignments thus far---see Table~\ref{tab:bug-table-summary}, and in \supp{} Table~\ref{tab:bug-table}.

    \item Importantly, we discovered QUIC specification ambiguities in connection management that expose a DoS vulnerability and propose an amendment to address the issue---see Section~\ref{sec:ambiguity}.
\end{itemize}

\vspace{-0.5mm}
\begin{table}[!h]
\caption{Tested QUIC implementations and their \textit{security} configurations. These include 
Google~(Google-quiche), Mozilla~(Neqo), Lite Speed Technologies (LSQUIC), Meta~(Mvfst), Microsoft (MsQuic), Cloudflare~(Quiche), Amazon~(S2n-quic) and Alibaba (XQUIC).}
\vspace{-0.5mm}
\label{tab:quicImplementations}
\begin{threeparttable}
\resizebox{\columnwidth}{!}{
\begin{tabular}{@{}llcl@{}}
\toprule
\textbf{Name} & \textbf{\begin{tabular}[c]{@{}c@{}}Commit \\ Version \end{tabular}} & \textbf{Tested Configurations} & \textbf{URL} \\ \midrule
Aioquic & 239f99b8 & Basic, Retry, PSK & \footnotesize \url{https://github.com/aiortc/aioquic}\\
\midrule
Google-quiche & 42dab6be & Basic, ClientAuth, PSK & \footnotesize \url{https://github.com/google/quiche}\\
\midrule
Kwik & 745fd4e2 & Basic, Retry, PSK & \footnotesize \url{https://bitbucket.org/pjtr/kwik/src/master/}\\
\midrule
LSQUIC & 1b113d19 & Basic, PSK & \footnotesize \url{https://github.com/litespeedtech/lsquic} \\
\midrule
MsQuic & 5c070cdc & \begin{tabular}[c]{@{}c@{}}Basic, Retry, ClientAuth,\\RetryClientAuth, PSK \end{tabular} & \footnotesize \url{https://github.com/microsoft/msquic} \\
% old msquic: ff3059b5e5e0669a8012fafa7b75ebaf59b684e4
\midrule
Mvfst & a76144e1 & \begin{tabular}[c]{@{}c@{}}Basic, ClientAuth, PSK\end{tabular} & 
 \footnotesize \url{https://github.com/facebook/mvfst} \\
\midrule
Neqo & aaabc1c1 & Basic, Retry & \footnotesize \url{https://github.com/mozilla/neqo}\\
\midrule
Ngtcp2 & f65399b5 & \begin{tabular}[c]{@{}c@{}}Basic, Retry, ClientAuth,\\RetryClientAuth, PSK \end{tabular} & \footnotesize \url{https://github.com/ngtcp2/ngtcp2} \\
\midrule
Picoquic & d2f01093 & \begin{tabular}[c]{@{}c@{}}Basic, Retry, ClientAuth,\\RetryClientAuth, PSK \end{tabular} & \footnotesize \url{https://github.com/private-octopus/picoquic} \\
\midrule
PQUIC & 841c8228 & Basic, \add{PSK} & \footnotesize \url{https://github.com/p-quic/pquic}\\
\midrule
Quant & 511d91c3 & Basic, Retry, PSK & \footnotesize \url{https://github.com/NTAP/quant} \\
\midrule
Quiche & 24a959ab & \begin{tabular}[c]{@{}c@{}}Basic, Retry, ClientAuth,\\RetryClientAuth, PSK \end{tabular} & \footnotesize \url{https://github.com/cloudflare/quiche} \\
\midrule
Quiche4j & ea5effce & \begin{tabular}[c]{@{}c@{}}Retry \end{tabular} & \footnotesize \url{https://github.com/kachayev/quiche4j} \\
\midrule
Quic-go & f78683ab & \begin{tabular}[c]{@{}c@{}}Basic, Retry, ClientAuth,\\RetryClientAuth, PSK \end{tabular} & \footnotesize \url{https://github.com/quic-go/quic-go} \\
\midrule
Quicly & d44cc8b2 & Basic, Retry, PSK & \footnotesize \url{https://github.com/h2o/quicly} \\
\midrule
Quinn & \begin{tabular}{@{}c@{}}4395b969\\e1e1e6e3\end{tabular} & \begin{tabular}[c]{@{}c@{}}Basic, Retry, ClientAuth,\\RetryClientAuth, PSK \end{tabular} & \footnotesize \url{https://github.com/quinn-rs/quinn} \\
\midrule
Quiwi & b7b5dadb & \begin{tabular}[c]{@{}c@{}}Basic, Retry, PSK \end{tabular} & 
 \footnotesize \url{https://github.com/goburrow/quic} \\
\midrule
S2n-quic & ec651875 & Basic, Retry & \footnotesize \url{https://github.com/aws/s2n-quic} \\
\midrule
XQUIC & 00f62288 & Basic, PSK & \footnotesize \url{https://github.com/alibaba/xquic} \\
\bottomrule
\end{tabular}
}
\vspace{0.5mm}
\begin{tablenotes}[para]
\small
\textbf{Basic}:~Basic handshake. \textbf{Retry}:~Handshake with client address validation. \\ \textbf{ClientAuth}:~Handshake with client authentication. \textbf{RetryClientAuth}:\\ Handshake with client address validation and authentication. \textbf{PSK}:~Hand-\\shake with pre-shared key.
\end{tablenotes}
\end{threeparttable}
\end{table}

\vspace{2mm}
\noindent\textbf{Responsible Disclosure.~}Following the practice of responsible disclosure, we shared our findings with corresponding development teams by sending bug reports to vendors/developers following their reporting policies. We summarize the current state of disclosures and vendor responses in Table~\ref{tab:bug-table} within the \supp{}.

\vspace{-1mm}
%-------------------------------------------------------------------------------
\section{Background} \label{sec:protocolStateFuzzing}
%-------------------------------------------------------------------------------
We provide a brief overview of the QUIC protocol handshake (secure connection establishment prior to application data exchange) and active automata learning 
add{before delving} into our framework in Section~\ref{sec:implementation}. 

\vspace{-2mm}
%-------------------------------------------------------------------------------
\subsection{QUIC Protocol Transport Layer Security} \label{sec:quicHandshake}
%-------------------------------------------------------------------------------
To understand the process of secure connection establishment in QUIC, we begin with a primer on connection establishment in a client-server setting. An entity using QUIC must complete a handshake with its endpoint before it can communicate. QUIC combines both transport and cryptographic parameter negotiations into a single handshake. Our work focuses on this QUIC handshake, which is responsible for establishing \emph{secure} multi-stream connections. 

QUIC provides \textbf{five} different security configurations, we briefly discuss each and the respective packets and frames employed. A simplified illustration of QUIC handshake message exchange for the 5 different secure connection configurations is shown in Figure~\ref{fig:quicHandshake}. These handshakes include:

\begin{figure}[b!]
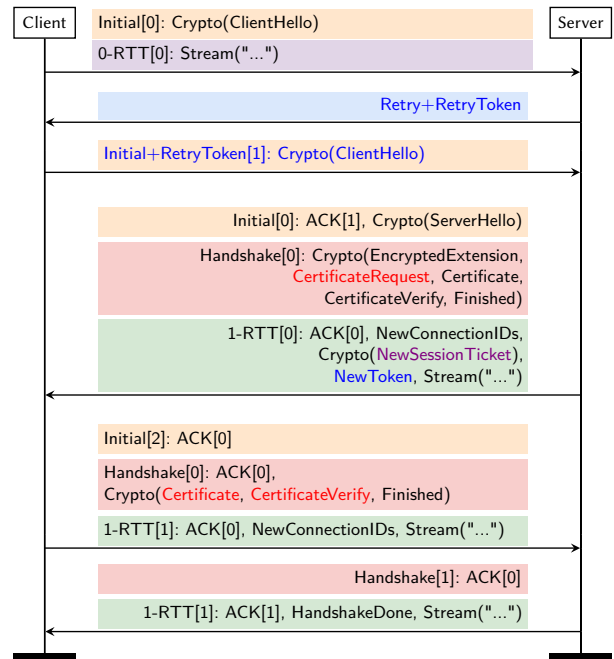

\vspace{-4mm}
%\resizebox{1\columnwidth}{!}{
\centering
 \resizebox{0.96\columnwidth}{!}{
    \begin{msc}[instance distance = 8.5cm, draw frame = none, environment distance = 0, instance width = 1.1cm, msc keyword=, head top distance=0cm]{} %
    \centering
    \declinst{client}{}{\textsf{Client}}
    \declinst{server}{}{\textsf{Server}}
    \mess[label distance = 0.3ex, align=left]{
    \colorbox{BackgroundOrange}{\parbox{7.7cm}{\raggedright \pmsg{{Initial[0]: Crypto(ClientHello)}}}} \\
    \colorbox{BackgroundPurple}{\parbox{7.7cm}{\raggedright \pmsg{{0-RTT[0]: Stream("...")}}}}}{client}{server}
    \nextlevel[1.8]
    \mess[label distance = 0.3ex, align=right]{\colorbox{BackgroundBlue}{\parbox{7.5cm}{\raggedleft \textcolor{blue}{\pmsg{{Retry+RetryToken}}}}}}{server}{client}
    \nextlevel[1.8]
    \mess[label distance = 0.3ex, align=left]{\colorbox{BackgroundOrange}{\parbox{7.5cm}{\textcolor{blue}{\pmsg{{Initial+RetryToken[1]: Crypto(ClientHello)}}}}}}{client}{server}
    \nextlevel[8]
    \mess[label distance = 0.3ex, align=right, draw]{
    \colorbox{BackgroundOrange}{\parbox{7.5cm}{\raggedleft \pmsg{{Initial[0]: ACK[1], Crypto(ServerHello)}}}}\\[0.5ex]
    \colorbox{BackgroundRed}{\parbox{7.5cm}{\raggedleft \pmsg{{Handshake[0]: 
    Crypto(EncryptedExtension,\\
    \textcolor{red}{CertificateRequest},
    Certificate,\\
    CertificateVerify,
    Finished)}}}}\\[0.5ex]
    \colorbox{BackgroundGreen}{\parbox{7.5cm}{\raggedleft \pmsg{{1-RTT[0]: ACK[0], NewConnectionIDs,\\
    Crypto(\textcolor{violet}{NewSessionTicket}),\\ \textcolor{blue}{NewToken}, Stream("...")}}}}
    }{server}{client}
    \nextlevel[5.5]
    \mess[label distance = 0.3ex, align=left]{
    \colorbox{BackgroundOrange}{\parbox{7.5cm}{\raggedright \pmsg{{Initial[2]: ACK[0]}}}}\\[0.5ex]
    \colorbox{BackgroundRed}{\parbox{7.5cm}{\raggedright \pmsg{{Handshake[0]: ACK[0], \\
    Crypto(\textcolor{red}{Certificate}, \textcolor{red}{CertificateVerify},
    Finished)}}}}\\[0.5ex]
    \colorbox{BackgroundGreen}{\parbox{7.5cm}{\raggedright \pmsg{{1-RTT[1]: ACK[0], NewConnectionIDs, Stream("...")}}}}
    }{client}{server}
    \nextlevel[3]
    \mess[label distance = 0.3ex, align=right]{
    \colorbox{BackgroundRed}{\parbox{7.5cm}{\raggedleft \pmsg{{Handshake[1]: ACK[0]}}}}\\[0.5ex]
    \colorbox{BackgroundGreen}{\parbox{7.5cm}{\raggedleft \pmsg{{1-RTT[1]: ACK[1], HandshakeDone, Stream("...")}}}}
    }{server}{client}
    \end{msc}
    }
    \vspace{-9mm}
    \caption{A \textit{simplified} overview of handshake security configurations consisting of \colorbox{BackgroundOrange}{Initial packets}, \colorbox{BackgroundPurple}{0-RTT packets}, \colorbox{BackgroundBlue}{Retry packets}, \colorbox{BackgroundRed}{Handshake packets} and \colorbox{BackgroundGreen}{1-RTT packets}. Frames \textcolor{blue}{for Address validation are in blue text} and for \textcolor{red}{Client Authentication are red text}. Messages in \textcolor{purple}{purple text carries the negotiated parameters to derive the pre-shared key for 0-RTT encryption in a future connection}. Packet space numbers for each packet type are within square brackets.}
    \label{fig:quicHandshake}
%}

\end{figure}

%%%%%%%%%%%%%%%%%%%%%%
\begin{enumerate}[itemsep=0pt,parsep=1pt,topsep=0pt]
    \item Basic.
    \item Client address validation (without client authentication).
    \item Client authentication and without address validation.
    \item Client address validation and authentication.
    \item Handshake with a pre-shared key.
\end{enumerate}

The handshakes are realized in QUIC using: \pmsg{Initial} packets, \pmsg{Handshake} packets, \pmsg{1-RTT} (Round Trip Time) packets and \pmsg{Retry} packets. These packets carry the necessary information in frames and QUIC messages to complete the transport and cryptographic parameter negotiations. Each frame is defined in~\cite[Section 12.4]{9000} to carry different types of data. For example, cryptographic parameters are in a CRYPTO frame. To simplify our explanations, we refer to frames used for connection establishment and the messages encapsulated within the frames such as CRYPTO frames as simply \textit{messages}.  

\vspace{2px}
\noindent\textbf{(1)~Basic Handshake.~}It is the most fundamental handshake~\cite[Section 7.1]{9000}. First, a client sends the server an \pmsg{Initial} packet with a \pmsg{ClientHello} message with application protocol negotiation, transport parameters, and cryptographic information to perform the key exchange. The server continues with an \pmsg{Initial} packet and a \pmsg{Handshake} packet. The \pmsg{Initial} packet contains a \pmsg{ServerHello} message containing cryptographic information to complete the key exchange. The \pmsg{Handshake} packet carries an \pmsg{EncryptedExtensions} message containing transport parameters and the negotiated application protocol version, a \pmsg{Certificate} message containing the server's certificate, a \pmsg{CertificateVerify} message used for requesting the client to verify the server's certificate, and a \pmsg{Finished message}. 

Notably, QUIC can combine multiple different types of packets into a single UDP datagram for transmission. Once the server sends the \pmsg{Finished} message, it can transmit application data using the Stream frame in \pmsg{1-RTT} packets. The client continues the handshake by verifying the server's certificate, and sending a \pmsg{Finished} message to the server. Both parties will verify the \pmsg{Finished} message received to ensure the previous handshake messages have not been modified.  After receiving a \pmsg{Finished} message from the client, the server will send a \pmsg{1-RTT} packet with a \pmsg{HandshakeDone} message to confirm the handshake and connection establishment.

\vspace{2px}
\noindent\textbf{(2)~Client Address Validation.~}Given that a QUIC server responds with many messages to a small initial request, it is at risk of exploitation for amplification attacks. QUIC provides an optional mechanism to validate a client's address, minimizing data sent to spoofed client IP addresses~\cite[Section 8]{9000}. Upon receiving the first \pmsg{ClientHello} message, the server responds with a \pmsg{Retry} packet containing a \pmsg{RetryToken}. A client that receives a \pmsg{Retry} packet must include the \pmsg{RetryToken} in all further \pmsg{Initial} packets sent for the rest of the handshake. The server will validate the \pmsg{RetryToken} contained in the client's subsequent \pmsg{Initial} packets. If the server fails to validate the \pmsg{RetryToken}, the server should immediately close the connection with a \pmsg{ConnectionClose} message. Moreover, after the server sends the \pmsg{Finished} message, the server can optionally send a \pmsg{NewToken} message with an address validation token that can be used for address validation in future connections.

\vspace{3px}
\noindent\textbf{(3)~Client Authentication.~}The server can select to authenticate a client by including a \pmsg{CertificateRequest} message in the \pmsg{Handshake} packet prior to sending the \pmsg{Finished} message~\cite[Section 4.4]{9001}. If the client receives a \pmsg{CertificateRequest} message, it must send a \pmsg{Certificate} message that contains the client's certificate and a \pmsg{CertificateVerify} message for client authentication. The server verifies the client certificate before sending the \pmsg{HandshakeDone} message. If the server fails to verify the client certificate, the server must close the connection with a \pmsg{ConnectionClose}.

\vspace{2px}
\noindent\textbf{(4)~Client Address Validation and Authentication.~}The complete handshake, using both client address validation and client authentication simultaneously, is illustrated in Figure~\ref{fig:quicHandshake}. The handshake incorporates the basic, client address validation and authentication protocol flow discussed above. 

\vspace{2px}
\noindent\textbf{(5)~Pre-shared key.~}A client can send \pmsg{0-RTT} packets carrying early (application) data to a server prior to handshake completion~\cite[Section 2.1]{9001}. The pre-shared key used to encrypt these packets is derived from the \pmsg{NewSessionTicket} message in the previous connection. This provisions for a faster, secure connection.

\vspace{2px}
\noindent\textbf{Encryption Keys and Packet Number Spaces.~}Instead of sequence numbers, QUIC uses three separate packet number spaces to track different packet types. \pmsg{Initial} packets use Initial packet number space; \pmsg{Handshake} packets use Handshake packet number space; \pmsg{0-RTT} and \pmsg{1-RTT} packets share Application data packet number space~\cite[Section 12.3]{9000}. Packets in different number spaces use different encryption keys~\cite[Section 4]{9001}. The \pmsg{Initial} packet uses the Initial key to provide the Initial encryption level with no confidentiality or integrity protection, the \pmsg{Handshake} packet uses the Handshake key in the Handshake encryption level, the \pmsg{0-RTT} packet uses the 0-RTT encryption key in 0-RTT encryption level, and the \pmsg{1-RTT} packet uses the 1-RTT encryption key in the 1-RTT encryption level. The Handshake, 0-RTT and 1-RTT encryption levels provide confidentiality and integrity.

%----------------------------------------------------------------------%---------
\vspace{-3mm}
\subsection{Automata Learning}
While automata learning can be categorized into passive and active learning, we focus on active learning~\cite{angluin1987learning, grinchtein2010learning}. In this setting, a learner constructs a deterministic finite automaton by generating queries to infer the behavior of a system by observing the resulting responses. Learner generates input sequences based on a dictionary of choices to probe a black-box system and observe the output symbol. To ensure a deterministic state model that precisely mirrors the behavior of a given black-box system, learning cycles through 2 phases: (i) \textit{hypothesis construction}; and (ii) \textit{conformance testing}. 

In the hypothesis construction phase, a series of input sequences and corresponding input-output observations are used to construct a hypothesis, a minimal deterministic state machine model that accurately reflects the recorded observations until this point for a possible FSM. The learner actively improves the hypothesis until conditions for convergence are fulfilled. For the input sequences that the learner had not sent and observed before, the hypothesis predicts an output by extrapolating from the recorded observations. To ensure this prediction accurately reflects the behavior of the black-box system, the learner proceeds to the conformance testing phase to validate the hypothesis. If the hypothesis is not supported by the behavior of the black-box system when a new input sequence is tested, the learner reverts back to the hypothesis construction phase to generate a more refined hypothesis. Alternatively, if the hypothesis matches the observed behavior of the black-box system for all the conformance tests, the learner considers it as the final learned state machine model, and the learning ends. Effectively, the learner treats the learned FSM as an equivalent oracle and searches for counterexamples to invalidate this assumption.

\vspace{-2mm}
%-------------------------------------------------------------------------------
\section{Noncompliance checking} \label{sec:implementation}
%-------------------------------------------------------------------------------
\pgfkeys{/csteps/inner color=white}
\pgfkeys{/csteps/fill color=black}
\begin{figure*}[t!]
    \centering
    \includegraphics[width=0.87\linewidth,trim={8mm 2mm 8mm 2mm},clip]{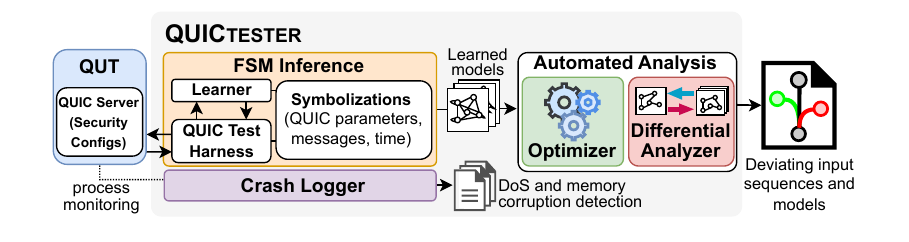}
    \vspace{-3mm}
    \caption{An overview of \name{}. FSM Inference with active learning and Automated Analysis method for identifying deviating behaviors.}
    \label{fig:QUIC-TeSterFramework}
    \vspace{-3mm}
\end{figure*}

In this section, we provide a high-level description of noncompliance checking framework and the design and implementation challenges along with our proposed solutions. 

We illustrate the \name{} framework modules in Figure~\ref{fig:QUIC-TeSterFramework}: i)~the \learner{}; ii)~\mapper{}; iii)~\opt{}; iv)~\logger{}; and  iv)~\differentialComparator. The \learner{} generates test inputs from symbolized QUIC protocol parameters, messages and event timing definitions. The \mapper{}, responsible for maintaining all protocol state with a test target, constructs the QUIC messages based on the symbolic instruction sequence for transmission through a UDP connection to the QUIC implementation under test (\sut{}). Subsequently, the \mapper{} awaits for the responses from the \sut{} within time intervals defined by the parameters in the input symbols. The \mapper{} deconstructs the received responses and reconstructs a set of output symbols using the symbolized messages to return to the \learner{}. The response symbols provide feedback to the \learner{} to explore and infer a FSM representation of the \sut{}--the \textit{learned model}.

We employ an \opt{} to simplify the complex models generated from the \learner{} to generate a simpler, easier to analyze automatically and more readable form of the learned model for analysis. Subsequently, the \differentialComparator{} automates the analysis process by using differential testing strategies to identify protocol non-compliance and potential security vulnerabilities. We discuss the design and implementation challenges (\textbf{C1}-\textbf{C7}) of the framework in the following.

\vspace{-3mm}
\subsection{Automatically Learning a Behavior Model (\learner{})}
\label{sec:Learner}

\noindent\Circled{C1}\textbf{~Generating test sequences (valid and invalid messages and parameters).} Fundamentally, in a blackbox setting, no prior knowledge is required. However, a \textit{protocol-specific symbolization} of possible inputs to combine and outputs for a target protocol must be defined to allow the \learner{} to generate tests---a series of input sequences. These sequences allow the effective exploration and inference of the FSM of the target system through not only valid but \textit{invalid} packet configurations as well as various security and transport parameter combinations in the learning phase. However, a symbolic dictionary for the purpose, extracted from a careful examination of the QUIC specification to test all 5 security settings in a QUIC handshake does not currently exist. Notably, symbolization is not straightforward. It demands a deep understanding of an extensive, technical specification to express and use protocol behaviors to facilitate uncovering an FSM and non-compliance.

\vspace{3px}
\noindent\textbf{Methods for resolving C1.~}To automatically define all the necessary symbols, we attempted to employ an LLM model to generate the symbols. However, the results returned by the LLM model were unsatisfactory---see results and discussion in Section~\ref{sec:llm_symbol_study_appendix} within the \supp{}---we reverted to manually extract the symbols from the specification. An author requires approximately 30-40 hours to read the specifications~\cite{9000,9001} and manually defines symbols for protocol parameters and messages for QUIC. Multiple researchers examined the RFC individually and agreed with all the symbol constructions, to avoid ambiguities. Although the construction phase is time-intensive, the symbolic dictionary needs to be constructed only once and can be applied in further studies on QUIC, thereby reducing the manual effort required to redefine the symbols. For constructing test inputs and modeling outputs, we include \totalsymb{} symbols. While Section~\ref{sec:quicHandshake} provides an overview of messages and their constituent parts, such as frames, we defer our justifications and details of symbolization to Appendix~\ref{sec:symb_appendix}.

\vspace{3px}
\noindent\Circled{C2}\textbf{~Learning Time-Dependent Behavior.~}We introduce the idea of testing with different time variations to uncover time-dependent vulnerabilities (such as \textbf{M}-4 in Table~\ref{tab:bug-table-summary}). However, using random timeouts is not beneficial as it can confuse the learner's observations and lead to non-determinism and failure to infer a FMS or complete the active learning process.

\vspace{3px}
\noindent\textbf{Methods for resolving C2.~}To observe time-dependent behavior, we parameterize the \textit{input} symbols. We determine two timeouts parameters, a $short$ timeout and a $long$, is adequate; as we explain below. The timeout parameter represent the duration the \mapper{} is required to wait prior to the subsequent message transmission after receiving a response; with $short$ representing the minimum wait time to receive \sut{}'s response for a request sent and $long$ is selected to be much longer, 10$\times$ $short$. The time parameterized symbols we curate are summarized in Table~\ref{tab:symbolDictionary}.

Importantly, the timeouts are discovered automatically by the \mapper{} prior to model learning. Notably, in contrast to an adversarial interpretation, the time parameter can also elicit behavior of the \sut{} under network delays experienced in practice. Two time settings are sufficient for capturing protocol states as they can capture results from when protocol implementation timeouts \textit{do} or \textit{do not} occur. Further timeout settings were found to increase learning time without leading to additional state discovery.

\vspace{3px}
\noindent\Circled{C3}\textbf{~Non-determinism during learning.~}Model learning depends on observing deterministic behavior from the \sut{} to infer a valid FSM model. We expect the \sut{} to generate deterministic responses to multiple repetitions of the same input sequence. However, since the \learner{} is unaware of time-related artifacts in the rest of the system, non-determinism can manifest when \textbf{(C3.1)}~the \mapper{} is not capable of receiving all the \sut{} responses during an allocated time period, for example, due to various execution speeds of QUIC implementations; and \textbf{(C3.2)}~the \learner{} sends an input before the \sut{} is fully initialized or \textbf{(C3.3)}~after the \sut{} crashes during learning, for example, if the \learner{} receives responses to a given input sequence in one step but fails to receive any when replaying the sequence because the \sut{} crashed before it can respond, the active learning task can fail. These sources of non-determinism can lead the \learner{} to infer an incorrect behavior or fail to complete the active learning task.

\vspace{2px}
\noindent\textbf{Methods for resolving C3.~}To address~\textbf{(C3.1)}, we adopt implemen\-tation-specific minimum time delays for the \mapper{} to wait to capture responses from the \sut{} to ensure that no responses are missed. Recall, the time for the \mapper{} to capture responses will be defined by the \learner{} as introduced in Section~\ref{sec:Learner}. Hence, we use the \mapper{} to record the longest time needed to capture expected packets in a handshake to determine the value for $short$, specific to each QUIC implementation, prior to model learning. For~\textbf{(C3.2)}, we augment a delay after starting the \sut{} and validate its availability before sending the first input. This ensures the \sut{} is initialized and ready to accept new connections. For~\textbf{(C3.3)}, we implement a \logger{} to monitor the status of the \sut{} after each test sequence iteration and restart the \sut{} if it crashes. These modifications ensure the \mapper{} and \learner{} behavior does not lead to non-deterministic outcomes and any observations of non-determinism are due to QUIC implementation defects.

\vspace{2px}
\noindent\textbf{\learner{} Implementation.~}We implemented the \learner{} using LearnLib, a Java library of active automata learning algorithms~\cite{learnlib}. We selected the TTT algorithm~\cite{isberner2014ttt}; it requires less queries compared to other algorithms~\cite{isberner2015foundations} and Wp-method algorithm~\cite{wpmethod} as the conformance testing algorithm.

\vspace{-4mm}
\subsection{QUIC Test Harness (\mapper{})}

\vspace{1mm}
\noindent\Circled{C4}\textbf{~Building a Test Harness to Support All Security Configurations, Valid and Invalid Messages, and Learning Time Dependent Behaviors.~}As described in Section~\ref{sec:protocolStateFuzzing}, the \mapper{} is a protocol-specific test harness responsible for exchanging messages with the \sut{}. It constructs and transmits QUIC messages sequentially, based on the test input symbol sequence determined by the \learner{}. The symbolic representation cannot be directly sent to a \sut{}. The \mapper{} must be functionally and logically correct on: \textbf{(C4.1)}~translating state and state transitions explored by the model \learner{} to QUIC protocol message exchanges with \sut{}; \textbf{(C4.2)}~constructing symbolic representations of protocol messages; whilst \textbf{(C4.3)}~maintaining state cohesion with the target \sut{} to achieve successful progression of a test from message to message. 
Further, as discussed in Section~\ref{sec:quicHandshake}, QUIC supports \textit{five} different secure connection establishments and requires deriving and installing 3 different encryption keys during a handshake. Therefore, building a \mapper{} is a significant undertaking. To the best of our knowledge, a QUIC-specific test harness to fulfill these requirements does not currently exist.

\vspace{2px}
\noindent\textbf{Methods for resolving C4.~}To manage the effort in addressing the challenges and building a comprehensive \textit{QUIC-specific test harness}, we extend and modify the Aioquic~\cite{aioquic} library. To address: 
\begin{itemize}[itemsep=1pt,parsep=1pt,topsep=1pt,labelindent=0pt,leftmargin=5mm]
    \item \textbf{(C4.1)}~we modify and extend the library functions to generate packets based on a given input symbol by the \learner{}. Further, to comprehensively test all \textit{five} different secure connection establishments, we implement functions for client authentication, which is currently \textit{not} supported by the library. 
    \item \textbf{(C4.2)}~we extend with functions to parse incoming packets from the \sut{} to output symbols. 
    \item \textbf{(C4.3)}~we utilize the state machine to maintain the state cohesion with the \sut{}. Importantly, we configured the state machine to not discard any installed encryption keys. This enables the \mapper{} to generate QUIC messages with the encryption key from the previous encryption level irrespective of the current encryption level. This capability is valuable for testing the server's ability to handle encryption keys across different encryption levels (Section~\ref{sec:quicHandshake}). For example, after the handshake is confirmed (1-RTT encryption level), the \mapper{} can still generate and transmit an \pmsg{Initial} packet (at the Initial encryption level) to the \sut{}. 
\end{itemize}

\vspace{-3mm}
\subsection{Automating Analysis (\opt{} \& \differentialComparator{})} \label{sec:differentialTesting}
\noindent\Circled{C5}\textbf{~Learned Models are Difficult to Interpret and Analyze.~}The learned model generated by the \learner{} is unnecessarily difficult to analyze due to artifacts from the testing phase and protocol complexity. 

\vspace{2px}
\noindent\textbf{Methods for resolving C5 (\opt{}).~}To ease the analysis process, we design an optimization routine to simplify the learned model by incorporating the following observation. The model description can contain numerous edges (state transitions) that do not provide useful information. In particular, the following types of edges observed in models are redundant: 

\begin{itemize}[itemsep=1pt,parsep=1pt,topsep=1pt,labelindent=0pt,leftmargin=5mm]
    \item Edges that do not transition to a different state.
    \item Edges with the same input but different timeouts transitioning to the same next state from a given state.
\end{itemize}

\noindent Therefore, we first, we remove all the edges that do not transition to a different state from the current state, as these edges indicate no progress on the \sut{}. Then, we merge all edges with the same input type but different timeouts, if they have the same current and next state. Doing so allows us to easily identify deviations given the same input with a different timeout. Including these optimizations in \name{} allowed up to 90\% of edges to be removed from the original learned model in the best case, while testing the Quinn server (Figure~\ref{fig:optimizedLearnedModel}) as shown in the \supp{}. The overall result is a considerable simplification to aid automated analysis and a significant reduction in the effort required to read and analyze models and to identify anomalous behavior.

\begin{figure}[t!]
    \centering
    \subfloat[Comparison between models originating
from the same target (server) and configuration but under different time parameter settings.]{\includegraphics[width=1\columnwidth]{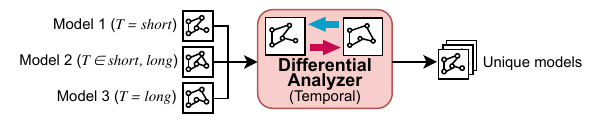}\label{fig:differentialTestingOnTemporal}}

    \vspace{-2mm}
    \subfloat[Fully automated testing method to mitigate the need for a reference model. Comparison between all unique models from Figure~\ref{fig:differentialTestingOnTemporal} to extract all input sequences that lead to deviations along with model descriptions (FSMs).]{\includegraphics[width=1\columnwidth]{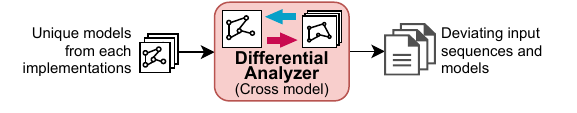}\label{fig:differentialTestingOnReferrence}}\\
    \vspace{-2mm}
    \caption{Automated analysis using \differentialComparator{}.}
     \label{fig:automatedAnalysis}
     \vspace{-3mm}
\end{figure}
\begin{figure}[t!]
    \centering
    \includegraphics[width=1\columnwidth]{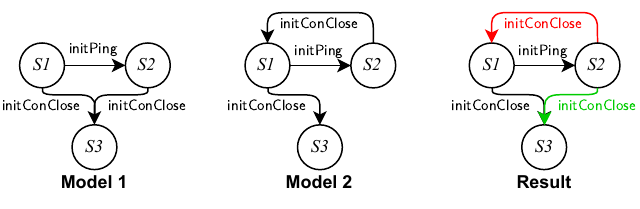}
    \vspace{-8mm}
    \caption{Example illustrating a \differentialComparator{} result. Given two optimized models as inputs (Model 1 and Model 2), unique state transitions (the green edge denotes the unique edge in Model 1 while the red edge denotes it is unique to Model 2) are identified by the algorithm.}
    \label{fig:lts_diff}
    \vspace{-3mm}
\end{figure}

\vspace{1mm}
\noindent\Circled{C6}\textbf{~Cost and Problems with Manual Deviation Analysis.~}One of the significant problems with black-box noncompliance checking using automata learning in the previous studies~\cite{tlsFuzzer,dtlsFuzzer} is the subsequent manual effort needed to analyze learned models by a domain expert. 

\noindent\textbf{Methods for resolving C6 (\differentialComparator{}).~}We develop an automated state machine comparison approach for the problem. A simplified example illustrating the automated analysis approach is shown in Figure~\ref{fig:lts_diff}. 

We prune all the output symbols from the optimized learned models and use the Labeled Transition System Differential (LTS\_Diff) algorithm~\cite{lts_diff} in two scenarios: i)~we compare models originating from the same target (server) and configuration but in different timeout settings (Figure~\ref{fig:differentialTestingOnTemporal}); and ii)~we cross check all unique models obtained from (i) against all other targets (Figure~\ref{fig:differentialTestingOnReferrence}) and extract \textit{all} the deviations automatically. During the comparison, we identify and highlight deviations---additional (green) or missing (red)---states and state transitions. Figure~\ref{fig:lts_diff} shows an example; the highlighted deviations indicate at least one of the implementations may not be compliant with the specification. To further analyze the deviations, we automatically extract the shortest input sequence with its corresponding output leading to deviations. To this end, we \textit{\textbf{only need to analyze the extracted deviations}} from (ii) to identify the non-compliant implementation. The approach significantly reduces the manual effort to inspect all the learned models. We discuss the reduced manual effort obtained from the approach in Section~\ref{sec:experiments}.

Significantly, our extensive testing regime has allowed us to curate conforming learned models to serve as reference models. The use of \name{} with the reference models can automatically extract non-conformance behaviors from the learned model we discuss in Section~\ref{sec:discussion}. Consequently, our automated approach in the \differentialComparator{} can support developers in the future by reducing the burden in identifying the non-compliant implementations.

\vspace{1mm}
\noindent\Circled{C7}\textbf{~Detecting DoS.~}If the \sut{} exhibits a memory-corruption bug or a logical flaw, unexpected behavior such as a crash can occur. In some cases, the crash can be hidden in the learned model~(as exemplified in \textbf{M}-8 in Table~\ref{tab:bug-table-summary}) and pose a challenge for detecting DoS vulnerabilities.

\vspace{1mm}
\noindent\textbf{Methods for resolving C7 (\logger{}).~}We employ a \logger{} to monitor the aliveness of a \sut{} by checking its \texttt{PID} after each learning step. If the \sut{} crashes, usually indicated by a missing or invalid \texttt{PID}, the \logger{} saves the \texttt{stderr} and \texttt{stdout} from the \sut{} with the corresponding \textit{input sequences} for subsequent analysis. The logging function and data allow identifying DoS attacks and memory-corruption bugs.

%-------------------------------------------------------------------------------
% \vspace{-1mm}
\section{Evaluation}\label{sec:experiments}
%-------------------------------------------------------------------------------
We tested and analyzed \implementationcount{} open-source QUIC implementations summarized in Table~\ref{tab:quicImplementations} using our \name{} implementation. Our implementation effort is summarized in Table~\ref{tab:code-lines} in Appendix~\ref{sec:impefffort_appendix}.

\vspace{1mm}
\noindent\textbf{Test Environment.~}All experiments are conducted using Ubuntu 20.04 with an AMD Ryzen 9 5950X CPU and 128~GB of RAM.

\vspace{1mm}
\noindent\textbf{QUIC Configurations.~} \label{sec:implementationsTestedAndAnalysed}
We test QUIC implementations with all mandatory and recommended cipher suites~\cite{TLS1.3RFC} implemented by the targets. Since all of the QUIC implementations we employed provide servers in their repository, we use the provided tools to configure and run these servers as the \sut{}. As mentioned in Section \ref{sec:quicHandshake}, our aim is to test all 5 different handshake configurations. Where an implementation did not support all 5 configurations, we tested the implemented configurations as summarized in Table~\ref{tab:quicImplementations}.

\vspace{1mm}
\noindent\textbf{Differential Testing With Time Parameters.~}In our experiments, we conduct learning on each implementation with three timeout settings: $T \in {short, long, mixed (short + long)}$. In our differential analysis, we compared the learned models generated from these three different timeout settings to identify state transitions that were influenced by temporal factors. 

\vspace{1mm}
\noindent\textbf{Identifying Anomalous Behaviors.~}We examined:
i)~differential test results with \name{} and validated the deviating behaviors discovered with the extracted input sequences to identify non-compliant behaviors; and ii) crashes, trace data and crashing seeds from the \name's crash logger. Subsequently, we examined the server code of these QUIC implementations for \textit{root cause analysis} and bug disclosures.

%%%%%% summary of faults
%% Bug-table %%
\begin{table}[t!]
%\vspace{-1mm}
\caption{Overview of identified faults (detailed in Table~\ref{tab:bug-table}).  
}
\label{tab:bug-table-summary}
\centering
\begin{threeparttable}
\resizebox{\columnwidth}{!}{%
\begin{tabular}{@{}p{3.5cm}p{6.5cm}c@{}}
\toprule
\textbf{Server} & \textbf{Fault Description} & \textbf{Type}-ID \\ \toprule
Aioquic & Incorrect handling of unexpected frame type. & \textbf{S}-1 \\ \midrule

Kwik & \href{https://nvd.nist.gov/vuln/detail/CVE-2024-22588}{Retention of the unused encryption keys.} & \textbf{S}-2 \\ %\cline{2-5} 
& \href{https://nvd.nist.gov/vuln/detail/CVE-2024-22590}{Implementation without a state machine.} & \textbf{S}-3 \\ 
& Process CRYPTO frame in a 0-RTT packet. & \textbf{S}-4  \\ 
& Exceeds the operating system’s maximum number of memory mappings for a single process (100,000) when receiving PING frame from 50,000 clients. & \textbf{M}-1 \\ \midrule

Lsquic~(Lite Speed) & \href{https://nvd.nist.gov/vuln/detail/CVE-2024-25678}{Retention of the unused encryption keys (PSK configuration).} & \textbf{S}-5 \\ 
 & Incorrect handling of re-transmission, leaving a half-opening connection on the client side (PSK configuration in v4.0.2). & \textbf{L}-1 \\ \midrule

MsQuic~(Microsoft) & Does not issue its initial\_source\_connection\_id at the correct connection state. & \textbf{S}-6\\ \midrule

Neqo~(Mozilla) & NULL pointer dereference when getting the primary path. & \textbf{M}-2 \\ 
 & Limited connections due to a hardcoded value. & \textbf{M}-3 \\ \midrule

Picoquic & NULL pointer dereference when getting the encryption keys. & \textbf{M}-4 \\ %\cline{2-5} 
 & Retry token tied to retry\_source\_connection\_id.  & \textbf{S}-7 \\ \midrule

PQUIC & Invalid original\_destination\_connection\_id. & \textbf{S}-8 \\ %\cline{2-4} 
 & Limitless active\_connection\_id\_limit. & \textbf{S}-9 \\ %\cline{2-4} 
& \href{https://nvd.nist.gov/vuln/detail/CVE-2024-25679}{Retention of the unused encryption keys.} & \textbf{S}-10\\
& Incorrect way of emptying the re-transmission queue. & \textbf{L}-2\\ 
& NULL pointer dereference when handling removed connection context. & \textbf{M}-5 \\ 
& Buffer overflow when processing frame type 0x30. & \textbf{M}-6 \\ 
& Infinite loop when processing frame type 0xFF. & \textbf{L}-3 \\
& Does not send HANDSHAKE\_DONE after the handshake is confirmed (PSK configuration). & \textbf{S}-11 \\ \midrule

Quiche~(Cloudflare) & Client authentication bypass due to incorrect flag set in Quiche library. & \textbf{S}-12 \\
 & Incorrect handling of \textbf{all} Initial packets carried in a UDP datagram with a payload size smaller than 1200 bytes. & \textbf{S}-13 \\ \midrule

Quiche4j & Concurrent modification exception when discarding closed connections. & \textbf{M}-7 \\ 
& Limitless active\_connection\_id\_limit. & \textbf{S}-14 \\
\midrule

Quant & Incorrect handling of an \pmsg{initialPing} message. & \textbf{S}-15 \\
 & Incorrect handling of \textbf{all} Initial packets carried in a UDP datagram with a payload size smaller than 1200 bytes. & \textbf{S}-16 \\  \midrule

Quiwi & Does not close the connection when the number of received NEW\_CONNECTION\_ID frames exceed the active\_connection\_id\_limit. & \textbf{S}-17 \\  \midrule

Quinn & \href{https://nvd.nist.gov/vuln/detail/CVE-2023-42805}{Panic when unwrapping a None value when processing an unexpected frame type.} & \textbf{M}-8 \\
 & Process CRYPTO frame in 0-RTT packet. & \textbf{S}-18 \\ \midrule
XQUIC~(Alibaba) & Retention of the unused encryption keys. & \textbf{S}-19 \\ %\cline{2-4} 
 & Maintaining a number of active connection IDs that
exceed the active\_connection\_id\_limit. & \textbf{S}-20 \\ \midrule

\begin{tabular}[c]{@{}l@{}}Aioquic, LSQUIC, Neqo, \\Quic-go, Quinn, Quiwi, S2n-\\quic~(Amazon), XQUIC \end{tabular}& \begin{tabular}[c]{@{}l@{}}Incorrect handling of the second and subsequent \\ Initial packets carried in a UDP datagram with a \\payload size smaller than 1200 bytes. \end{tabular} &  \begin{tabular}[c]{@{}c@{}}\textbf{S}-21 \\to \\ \textbf{S}-28\end{tabular}  \\ \midrule

\begin{tabular}[c]{@{}l@{}}Aioquic, Kwik, MsQuic, LS-\\Quic, Quant, Quiche, Quic-\\go, Quiche4j, Quiwi, S2n-quic \\ \end{tabular} & \begin{tabular}[c]{@{}l@{}}Accept Handshake packet from an unmatched \\ Destination Connection ID. \\ \\ \end{tabular} & \begin{tabular}[c]{@{}c@{}}\textbf{S}-29 \\to \\\textbf{S}-38 \end{tabular}  \\ \midrule

\begin{tabular}[c]{@{}l@{}}Lsquic, MsQuic, Neqo, \\Quiche4j, Quinn, XQUIC \\ \\ \end{tabular} & \begin{tabular}[c]{@{}l@{}}Incorrect handling of packets without a frame. \\ \\ \\ \end{tabular} &  \begin{tabular}[c]{@{}c@{}}\textbf{S}-39 \\to \\ \textbf{S}-44 \end{tabular}\\
\bottomrule
\end{tabular}
}

\begin{tablenotes}
    \item {\footnotesize In total \protocolViolation{} specification violations (\textbf{S}),
    \softwareBugs{} memory-corruption bugs (\textbf{M}) and \logicBugs{} logical }
    \item {\footnotesize flaws (\textbf{L}) were identified across \implementationcount{} implementations.}
  \end{tablenotes}
\end{threeparttable}
\vspace{-5mm}
\end{table}

%-------------------------------------------------------------------------------
\vspace{-1mm}
\section{Results and Analysis} \label{sec:stateMachineAnalysis}
%-------------------------------------------------------------------------------
%{\ttfamily\hyphenchar\font=`\-}
In this section, we discuss our results from analyzing the \modelcount{} learned models. With the model optimization technique, our framework can generate a more readable learned model for analysis. To simplify the presentation of learned models, in the following analysis: i)~we group the input and output symbols on a transition to a connection close, and label it with the \pmsg{{Other}} symbol; and ii)~when all input and output symbols from a given state transition to the same next state, we also group these symbols, replacing them with \pmsg{{Other}}, as these transitions do not provide any new information. Further, we have highlighted valid paths to complete a handshake in blue and highlighted invalid paths denoting adverse behavior in red. An example to illustrate the interpretation of a learned model is given in Appendix~\ref{appendix:valid_demo}. Here, we present a series of case studies on fault discoveries impacting the security or availability of servers.

\vspace{2px}
\noindent\textbf{Threat Model.~}Notably, our analysis is based on the threat model described in \cite{9000} and \cite{RFC3552}; wherein a key aspect of QUIC is to build mechanisms to mitigate DoS attacks. In the following, we present a series of summarized case studies on fault discoveries impacting the security or availability of servers. Detailed discussions of the case studies and a demonstration of valid behavior analysis on one of the reference models are included in the \supp{}.

% \vspace{-2mm}
\noindent\textbf{Results Summary.~}We summarize all the observed faults in Table~\ref{tab:bug-table-summary}; a detailed table with extended discussions is in Table~\ref{tab:bug-table} within the \supp{}. Each fault is categorized as one of the following: i)~\textit{specification bugs.} An implemented behavior violates the QUIC specification; ii)~\textit{memory-corruption bugs.} An input causing memory corruption and a server crash. iii)~\textit{logical flaws.} Incorrect logic implemented in code produces unexpected behavior.

\vspace{-2mm}
\subsection{Non-Compliance Issues}
\begin{mdframed}[backgroundcolor=black!5,rightline=false,leftline=false,topline=false,bottomline=false,roundcorner=1mm,everyline=false]
We discovered \protocolViolation{} specification violations---i.e. non-compliance issues. We present four case studies of significant issues and defer details and inputs for reproducing all of the issues to our GitHub repository~\cite{repo-inputs}.
\end{mdframed}

\label{sec:QuicheAppendix}

\begin{figure}[t!]
    \centering
    \includegraphics[width=\linewidth]{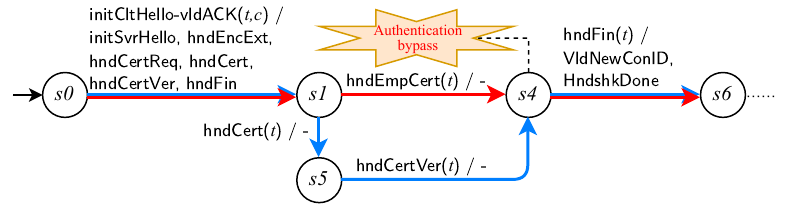}
    \vspace{-7mm}
    \caption{Simplified learned model of a Quiche server with the \clientAuthentication{} configuration. Blue edges show a valid path to complete a QUIC handshake. Red edges demonstrate an invalid path that bypasses the client authentication. The complete model is in Figure~\ref{fig:quicheClienttAuth}, in the \textit{Appendix}.}
    \label{fig:quicheClienttAuthSimple}
    \vspace{-5mm}
\end{figure}

\noindent\Circled{S-18} \textbf{Client Authentication Bypass in Quiche.}~The simplified Quiche learned model with \clientAuthentication{} configuration is shown in Figure~\ref{fig:quicheClienttAuthSimple}. The valid path to establish a QUIC handshake is highlighted in blue color. Our analysis of this model revealed that the server bypasses the client authentication on the path:~\emph{s0}, \emph{s1}, \emph{s4}, \emph{s6}, highlighted in red. This behavior was observed in all 3 handshake processes initiated with a different cipher suite($AES\_128, AES\_256$, $ChaCha20$). In this path, after the exchange of transport parameters and cryptographic information between the client and server, the server sends a \pmsg{{handshakeCertificateRequest}} message to authenticate the client. However, instead of responding with a valid certificate, the client sends a \pmsg{{handshakeEmptyCertificate}} message to the server at \emph{s1}. Subsequently, the client sends a \pmsg{{handshakeFinished}} message to complete the handshake. The server processes the client's \pmsg{{handshakeFinished}} message without verifying the client's certificate. Then, the server responds with \pmsg{{ValidNewConnectionID}} and \pmsg{{HandshakeDone}} messages to confirm the completion of a successful handshake. The summarized flow of this authentication bypass is illustrated in Figure~\ref{fig:quicheClientAuthBypass} in the Appendix.

During our code analysis of Quiche, we discovered the developers had set the incorrect flag ({\ttfamily\hyphenchar\font=`\- SSL\_VERIFY\_PEER}) for verifying the client certificate in the BoringSSL~\cite{boringssl} configuration. To address this issue, the flag should be set to {\ttfamily\hyphenchar\font=`\- SSL\_VERIFY\_PEER $|$ SSL\_VERIFY\_FAIL\_IF\_NO\_CERT}, which ensures that the handshake fails if an empty client certificate is used for authentication. However, according to BoringSSL documentation, this misconfiguration only affects the client authentication, i.e., an anonymous server will always fail to establish a connection with the client even without setting the {\ttfamily\hyphenchar\font=`\- SSL\_VERIFY\_FAIL\_IF\_NO\_CERT} flag.

\vspace{1mm}
\noindent\textit{\Circled{Impact}~}This allows an unauthenticated client to set up an anonymous connection with the server, allowing an on-path active attacker to perform man-in-the-middle attacks~\cite{TLS1.3RFC,TLS1.2RFC}. Notably, the issue was fixed with a bug bounty awarded to our team).

\noindent\Circled{S-19} \textbf{Retention of the unused encryption keys.}~We discover that the XQUIC server will still respond to \pmsg{Initial} packets after moving to the Handshake encryption level, as well as responding to \pmsg{Handshake} packets after moving to the 1-RTT encryption level (when the handshake is confirmed). XQUIC does not follow the specification as stated in~\cite[Section 4.9]{9001}, a QUIC server must discard the unused keys after moving to a new encryption level. For example, a server must discard its Initial key after it processes the first \pmsg{Handshake} packet from the client so that the subsequent \pmsg{Initial} packets will not be processed. 

The non-conformance described above can lead to serious consequences. \textit{For example}, an attacker can disrupt a connection. The Initial key does not provide confidentiality or integrity protection against attackers that can observe packets~\cite[Section 17.2.2]{9000}. Notably, the Initial key is determined by using HKDF-Extract with a default salt specified in~\cite[Section 5.2]{9001} and an input keying material (IKM) of the Destination Connection ID from the client's first \pmsg{Initial} packet. An attacker can sniff a victim's (client) first \pmsg{Initial} packet, obtain the Destination Connection ID and compute the victim's Initial key. Then, the attacker can use the victim's Initial key to send a spoofed \pmsg{initialConnectionClose} message to the server. \textit{A server that does not discard the Initial key may use this key for decryption, process the spoofed \pmsg{Initial} packet and} \textbf{\textit{close the connection with the victim}} (DoS attack).

\vspace{2px}
\noindent\textit{\Circled{Impact}~}The non-conformance behavior~\cite[Section 4.9]{9001} allows an off-path active attacker to disrupt a connection during the handshake in both security settings supported by XQUIC to mount a DoS attack. The attack requires a malicious actor able to sniff the first \pmsg{Initial} packet sent by the victim (a QUIC client) on a network.

%%%%%%%%%%%%%% Kwik %%%%%%%%%%%%%%

\vspace{1mm}
\noindent \Circled{S-2} \textbf{Retention of the unused encryption keys.~} Once a key is created for an encryption level, the Kwik server will continue decrypting and processing packets from that encryption level, even after moving to a new encryption level. Similar to XQUIC, this behavior is not conforming to~\cite[Section 4.9]{9001}. So, Kwik is also vulnerable to the spoofed \pmsg{initialConnectionClose} attack explained earlier.

\vspace{1mm}
\noindent\textit{\Circled{Impact}~}An off-path active attacker can mount a DoS attack by disrupting a connection in all security settings supported by Kwik. The attack requires a malicious actor able to sniff the first \pmsg{Initial} packet sent by the victim (a QUIC client). This vulnerability was assigned \textsf{CVE-2024-22588} and patched by the Kwik developers.

Notably, Kwik presents a more critical issue than XQUIC. Beyond just retaining the unused encryption keys, Kwik does not actually track the current state of a connection. In other words, Kwik does not implement a proper state machine---as discussed in \Circled{S-3} below.

\vspace{2px}
\noindent \Circled{S-3} \textbf{Implementation without a TLS state machine.~}In testing, we found the Kwik implementation to reprocess a message that was already successfully processed before, such as an \pmsg{{initialClientHello}} message. Notably, this message contains application protocol negotiation, transport parameters, and cryptographic information to perform key exchange. This reprocessing of an \pmsg{{initialClientHello}} message will overwrite the existing connection's application protocol version, transport parameters, and encryption key.

This is a serious issue because, combined with the vulnerability we discussed earlier, attackers can reset or potentially hijack a victim's connection using a \pmsg{{initialClientHello}} message. For example, at any state of a connection, an attacker with the victim's Initial key can send a spoofed \pmsg{{initialClientHello}} message with transport parameters and cryptographic information that differs from the victim's to the Kwik server. The Kwik server will process the spoofed \pmsg{{initialClientHello}} message and overwrite the existing transport parameters and encryption key that it has with the victim. Due to the desynchronization of transport parameters and encryption keys, the server no longer recognizes the victim and drops any packets coming from the victim. An attacker can then sniff the responses from the server, complete the overwritten handshake and use the connection to exchange data with the server.

\vspace{1mm}
\noindent\textit{\Circled{Impact}~}A malicious off-path active actor can hijack a victim's active connection. Notably, this vulnerability has been fixed by the developers with \textsf{CVE-2024-22590} assigned.

%%%%%%%%%%%%%%%%%%%%%%%%% Picoquic seg. fault  start &&&&&&&&&&&&&&&&&&&&&&&&&&&&&&&&&&&&
\vspace{-1mm}
\subsection{Memory-corruption bug: Server crashes} \label{sec:Picoquic}
\begin{mdframed}[backgroundcolor=black!5,rightline=false,leftline=false,topline=false,bottomline=false,roundcorner=1mm,everyline=false]
We discovered \softwareBugs{} memory-corruptions. We defer details and inputs for reproducing bugs to~\cite{repo-inputs}.  Here, we review M-4 uncovered with our \textit{idea for discovering timing dependencies on protocol states}. We include three further case studies in Appendix~\ref{appendix:memory_case_studies}.
\end{mdframed}

\label{sec:picoquic_memory}

\begin{figure}[t!]

    \centering
    \subfloat[Simplified model from Picoquic with \clientAuthentication{} configuration learned with $T=short$ parameter setting for inputs.]{\includegraphics[width=1\columnwidth]{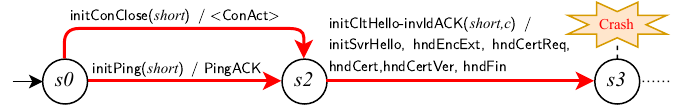}\label{fig:picoquicBWCA_simplified_short}}\\
    
    \subfloat[Simplified model from Picoquic with \clientAuthentication{} configuration learned with $T=long$ parameter setting for inputs.]{\includegraphics[width=1\columnwidth]{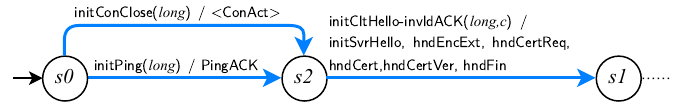}\label{fig:picoquicBWCA_simplified_long}}\\
    \vspace{-2mm}
\caption{Differential analyzer reveals a deviation on Picoquic when different time parameters are used for inputs. The complete models are given in Figure~\ref{fig:picoOptimisedShort} and~\ref{fig:picoOptimisedLong} in the \textit{Appendix}.}
\vspace{-5mm}
     \label{fig:picoquicBWCA_simplified}
\end{figure}

\noindent\Circled{M-4}~\textbf{Null Pointer Dereference in Picoquic.}~Interestingly, our use of differential testing with time parameters was crucial to identifying the issue. Figure \ref{fig:picoquicBWCA_simplified} depicts the Picoquic's \clientAuthentication{} simplified learned models using different time parameters. Our \differentialComparator{} revealed a deviating state transition from \emph{s2} in Figure~\ref{fig:picoquicBWCA_simplified_short} and Figure~\ref{fig:picoquicBWCA_simplified_long} when the same input, \pmsg{initialClientHello-invldACK} was sent. The state transitions in both figures are responded to with the necessary messages to continue the handshake. However, when a client sends any further messages at \emph{s3} in Figure~\ref{fig:picoquicBWCA_simplified_short}, the server does not respond. Data from our \logger{} revealed a segmentation fault occurred on each occasion the specific input sequence was received by the server at \emph{s3} in Figure~\ref{fig:picoquicBWCA_simplified_short}.

We use \texttt{rr}~\cite{rr} to investigate further and record Picoquic's execution while sending the crashing input sequence. We debug the root cause of this crash by replaying the recorded program execution. The segmentation fault error occurs because the server tries to re-transmit an \pmsg{{initialServerHello}} message and attempts to dereference a null pointer when retrieving the required encryption key.

\begin{figure}[t]
    \centering
    \includegraphics[width=0.55\linewidth, trim={2mm 1mm 2mm 2mm}]{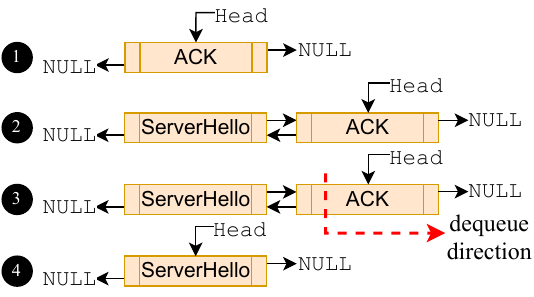}
    \vspace{-2mm}
    \caption{Changes in Picoquic's re-transmission queue stored as a double-linked list, leading to a segmentation fault.}
    \label{fig:picoquicDoubleLinkedList}
    \vspace{-3mm}
\end{figure}

\begin{figure}[t!]
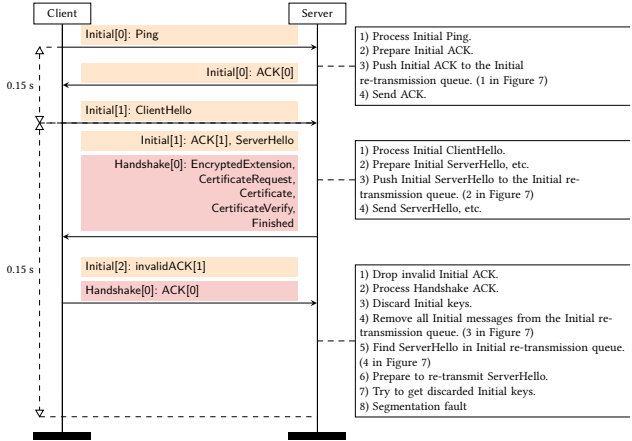

%\vspace{-2mm}
 \resizebox{1.0\columnwidth}{!}{
    \begin{msc}[instance distance = 5.2cm, draw frame = none, environment distance = 0, instance width = 1.5cm, msc keyword=, head top distance = 0cm]{}
    \centering
    \declinst{client}{}{\textsf{Client}}
    \declinst{server}{}{\textsf{Server}}
    \mess[label distance = 0.3ex]{\colorbox{BackgroundOrange}{\parbox{5.5cm}{\raggedright \pmsg{{Initial[0]: Ping}}}}}{client}{server}
    \measure{0.15~s}{client}{server}[4]
    \nextlevel[1]
   \msccomment[%
  msccomment distance=1cm,
  side=right,
  /msc/every msccomment/.append style={
   text width=7cm
  }
  ]{1) Process Initial Ping.\\ 
  2) Prepare Initial ACK.\\
  3) Push Initial ACK to the Initial \\re-transmission queue. (1 in Figure \ref{fig:picoquicDoubleLinkedList})\\
  4) Send ACK.}{server}
    \nextlevel[1]
    \mess[label distance = 0.3ex]{\colorbox{BackgroundOrange}{\parbox{5.5cm}{\raggedleft \pmsg{{Initial[0]: ACK[0]}}}}}{server}{client}
    \nextlevel[2]
    \mess[label distance = 0.3ex]{\colorbox{BackgroundOrange}{\parbox{5.5cm}{\raggedright \pmsg{{Initial[1]: ClientHello}}}}}{client}{server}
    \measure{0.15~s}{client}{server}[15.5]
    \nextlevel[3]
    \msccomment[%
  msccomment distance=1cm,
  side=right,
  /msc/every msccomment/.append style={
   text width=7cm
  }
  ]{1) Process Initial ClientHello.\\
  2) Prepare Initial ServerHello, etc.\\
  3) Push Initial ServerHello to the Initial re-transmission queue. (2 in Figure \ref{fig:picoquicDoubleLinkedList})\\
  4) Send ServerHello, etc.}{server}
    \nextlevel[3]
    \mess[label distance = -2ex, align=right]{\colorbox{BackgroundOrange}{\parbox{5.5cm}{\raggedleft \pmsg{{Initial[1]: ACK[1], ServerHello}}}}\\[0.5ex]
    \colorbox{BackgroundRed}{\parbox{5.5cm}{\raggedleft \pmsg{{Handshake[0]: EncryptedExtension,\\ 
    CertificateRequest,\\ 
    Certificate,\\ 
    CertificateVerify,\\ 
    Finished\\}}}}\\\\
    }{server}{client}
    \nextlevel[3.5]
    \mess[label distance = -2ex, align=right]{\colorbox{BackgroundOrange}{\parbox{5.5cm}{\raggedright \pmsg{{Initial[2]: invalidACK[1]}}}}\\[0.5ex]
    \colorbox{BackgroundRed}{\parbox{5.5cm}{\raggedright \pmsg{{Handshake[0]: ACK[0]}}}}\\[0.8ex]
    }{client}{server}
    \nextlevel[2]
    \msccomment[%
  msccomment distance=1cm,
  side=right,
  /msc/every msccomment/.append style={
   text width=7cm
  }
  ]{1) Drop invalid Initial ACK.\\
  2) Process Handshake ACK.\\
  3) Discard Initial keys.\\
  4) Remove all Initial messages from the Initial re-transmission queue. (3 in Figure \ref{fig:picoquicDoubleLinkedList})\\
  5) Find ServerHello in Initial re-transmission queue. (4 in Figure \ref{fig:picoquicDoubleLinkedList})\\
  6) Prepare to re-transmit ServerHello.\\
  7) Try to get discarded Initial keys.\\
  8) Segmentation fault}{server}.
  \nextlevel[3]
    
    \end{msc}
    }
    \vspace{-4mm}
    \caption{A message sequence chart showing message flow required to trigger a segmentation fault in Picoquic with the \clientAuthentication{} configuration.}
    \vspace{-4mm}
    \label{fig:picoquicSegFaultFlow}
\end{figure}

This segmentation fault error in Picoquic depends on several events within the server's operation. When the server acknowledges the \pmsg{initialPing}, it pushes its first message (\pmsg{PingACK}) to the head node of its Initial re-transmission queue in \circled{1} shown in Figure~\ref{fig:picoquicDoubleLinkedList}. Notably, the server uses a double-linked list as its re-transmission queue. In response to the client's \pmsg{{initialClientHello-invalidACK}}, the server sends its second message (\pmsg{{initialServerHello}}). The message is added to the previous node of the head node, as illustrated in \circled{2} in Figure~\ref{fig:picoquicDoubleLinkedList}. Subsequently, the server drops the invalid Initial ACK that acknowledges the \pmsg{{initialServerHello}} message and processes the \pmsg{Handshake} ACK that acknowledges the first \pmsg{Handshake} message it sent. As a result, the \pmsg{{PingACK}} and \pmsg{{initialServerHello}} remain in the Initial re-transmission queue. After processing the \pmsg{Handshake} ACK, the server removes its Initial keys as described in~\cite[Section 4.9.1]{9001}. Subsequently, the server attempts to remove all the messages in the Initial re-transmission queue to prevent any transmission of \pmsg{Initial} messages.

Interestingly, the server's method of emptying the Initial re-transmission queue does not follow the order in which the messages were added. The server removes the head node and all its sub-sequence messages stored in its next node as shown in \circled{3} in Figure~\ref{fig:picoquicDoubleLinkedList}. Consequently, only the oldest message is removed, while the remaining messages in the re-transmission queue remain as shown in \circled{4}. Later, when a re-transmission callback occurs, the server attempts to encrypt the \pmsg{{initialServerHello}} message by dereferencing the null pointer that previously stored the discarded Initial keys. This is the source of the segmentation fault error. The summarized flow of QUIC handshake protocol messages leading to the segmentation fault is illustrated in Figure~\ref{fig:picoquicSegFaultFlow} in the \textit{Appendix}.

Notably, when the first input, \pmsg{{initialPing}}, is set to $long$ timeout, the server removes the \pmsg{{PingACK}} from the Initial re-transmission queue before processing the \pmsg{Handshake} ACK. As a result, the \pmsg{{initialServerHello}} message becomes the head node in the queue. When the server discards the Initial keys, it also removes the first message in the queue, which is the \pmsg{{initialServerHello}} message. This proactive step prevents the server from encountering a segmentation fault and allows it to complete the handshake successfully.

Interestingly, the complete \clientAuthentication{} model learned with inputs ($T=short$) has 11 states and shows the adverse behavior described above, while the complete \clientAuthentication{} model learned with inputs ($T=long$) has only 9 states with no anomalies (complete learned models can be found at~\cite{repo-inputs}). As evidenced, the difference in models demonstrates testing with time-parameterized inputs elicits new behavior of a network protocol implementation, leading to new states, state transitions, and potential new bug discoveries. 

\vspace{2mm}
\noindent\textit{\Circled{Impact}~}The vulnerability allows an attacker to perform a simple DoS attack with a single client during connection establishment.

\vspace{-2mm}
\subsection{Logical Flaw: Unexpected Behavior} \label{sec:logical-bug}
\begin{mdframed}[backgroundcolor=black!5,rightline=false,leftline=false,topline=false,bottomline=false,roundcorner=1mm,everyline=false]
Logical flaws produce unexpected behavior due to incorrect program logic. We discovered \logicBugs{} logical flaws but defer details and inputs for reproducing all issues to our GitHub repository~\cite{repo-inputs}. We review L-1 discovered by \name{} here and include two more in Appendix~\ref{appendix:bug_case_studies}.
\end{mdframed}

\vspace{2mm}
\noindent\Circled{L-1} \textbf{Incorrect handling of re-transmissions.}~During connection establishment, Lsquic server with \psk{} configuration will reach the close state every time it attempts to re-transmit the last unacknowledged message. 

\vspace{1mm}
\noindent\textit{\Circled{Impact}~}A client unable to acknowledge the server's handshake messages in time will always fail to establish a connection with the server.

\vspace{-2mm}
\subsection{Specification Ambiguity: Exposing a New DoS Attack} 
\label{sec:ambiguity}
When comparing the optimized models across \implementationcount{} server implementations, we identified a scenario under which an ambiguity related to the connection management aspects in the QUIC specification can lead to implementations exhibiting different behaviors; some more detrimental than others.

\vspace{1mm}
\noindent The specification states:

\vspace{1mm}
\parbox{0.90\columnwidth}{\texttt{The first packet sent by a client always includes a CRYPTO frame that contains the start or all of the first cryptographic handshake message.} (Section 17.2.2, RFC 9000).}
\vspace{2mm}

However, the specification does not explicitly state how a server should/must handle a first packet received without a CRYPTO frame nor what first packets a server must process. Further, as described in the specification:

\vspace{1mm}
\parbox{0.90\columnwidth}{\texttt{After processing the first Initial packet, each endpoint sets the Destination Connection ID field in subsequent packets it sends to the value of the Source Connection ID field that it received.} \\(Section 7.2, RFC 9000),}
\vspace{2mm}

\noindent A server will also need to ``remember'' the Connection IDs (or connection context) sent from clients after processing the first packet, \textit{even if the client has no intention of initiating a connection with the server}. When a client sends an \pmsg{initialPing} as the first packet to assess reachability of a server, we observed:

\begin{itemize}
[itemsep=2pt,parsep=1pt,topsep=1pt,labelindent=0pt,leftmargin=5mm]
    \item A few implementations elect not to acknowledge pings from clients that do not have an existing open connection with the server. These servers elect to drop the packet, or respond with a \pmsg{connectionClose} message.

    \item However, many implementations accepting \pmsg{initialPing} packets from clients \textit{without} an existing connection, create a connection based on the ping and respond with an acknowledgment. In this behavior, the server is forced to ``remember'' the connection ID (leading to creating a \textit{connection context}) \textit{even if the client has no intention of initiating a connection with the server}. Hence, irrespective of the client continuing the handshake, the server has expended resources in creating a connection context. These resources are not de-allocated until a timeout occurs.

    \item Only one implementation (MsQuic) responds to the ping without creating a connection context. 

\end{itemize}

\begin{figure}[h!]
 \resizebox{1\columnwidth}{!}{
    \centering
     \vspace{-7mm}
    \includegraphics[width=1\textwidth,trim={0mm 2mm 0mm 5mm},clip]{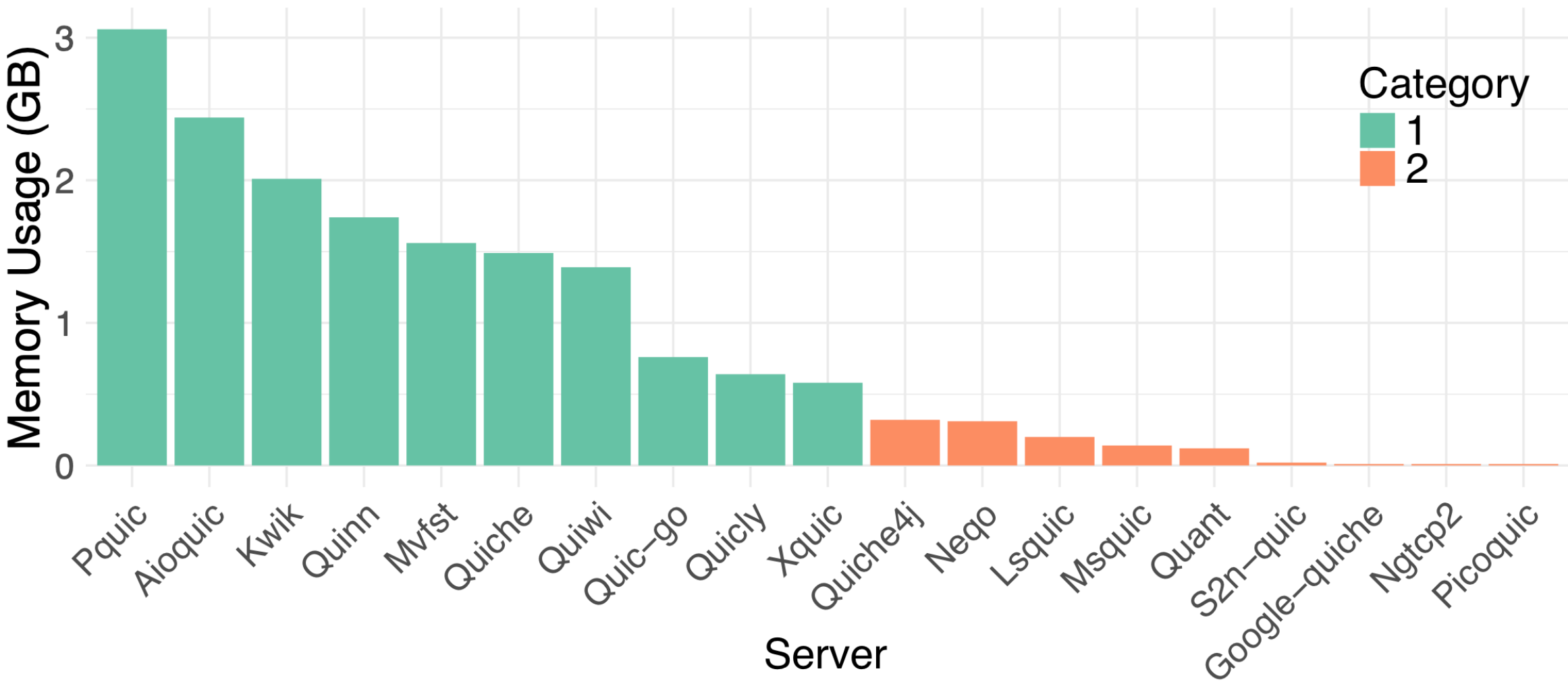}}
    \caption{Memory usage recorded for \implementationcount{} QUIC servers tested with 50,000 clients sending an \pmsg{Initial} packet with a PING frame without a CRYPTO frame.}
    \label{fig:maxMemoryUsage}
    \vspace{-4mm}
\end{figure}

To investigate the impact of implementation choices, we experiment by initiating 50,000 QUIC clients. Each client sends one packet, \pmsg{initialPing}, to a target QUIC server with \textbf{Basic} configuration. Since the specification states the first packet always contains a CRYPTO frame and servers can elect to drop the \pmsg{initialPing} without a CRYPTO frame, we tested with \pmsg{initialPing} packets without a CRYPTO frame. We tested the \implementationcount{} QUIC servers we studied. During the experiment, we made an interesting observation---a significant increase in memory usage (from 500~MB to 3~GB) for 10 server implementations, denoted as \textit{Category 1} in Figure~\ref{fig:maxMemoryUsage}. This is a direct consequence of the servers always creating a connection context for each incoming \textit{first} packet, even when a client does not intend to establish a connection (\pmsg{initialPing} that excludes a CRYPTO frame).

Interestingly, we were able to \textbf{\textit{crash the Kwik server}} \textbf{(M}-1\textbf{)}   \textbf{\textit{after it exceeded the operating system’s maximum number of memory mappings}} for a single process (100,000) (the server created 2 threads for each incoming \pmsg{initialPing}). We disclosed our findings to each affected QUIC implementation developer. 

We propose the specification to allow responding to clients not seeking to establish a connection to support liveness testing via \pmsg{initialPing} without a CRYPTO frame where a connection context is not created and amending the specification to state a first packet to initiate a connection \textit{MUST} include a CRYPTO frame.

\vspace{-2px}
\begin{figure}[b!]
\vspace{-3mm}
    \centering
    \includegraphics[width=0.47\columnwidth,trim={7mm 7mm 6mm 6mm},clip]{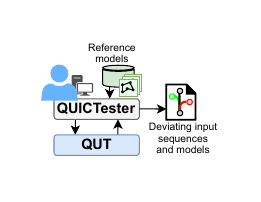}
    \vspace{-3mm}
    \caption{Practitioners can use \name{} with our curated reference models to automate future testing of target QUIC implementations (QUT).
    }
    \label{fig:QUIC-TeSter-use-case}
\end{figure}

%%%%%%%%%%%%%%%%%%%% Comparison and Fig 8 %%%%%%%%%%%%%%%%%%
\begin{figure*}[h!]
\begin{minipage}{1\textwidth}
\begin{table}[H]
\setlength{\extrarowheight}{0pt}
\setlength{\aboverulesep}{0pt}
\setlength{\belowrulesep}{0pt}
\centering
\caption{Comparison of automated testing studies on QUIC (Both Compliance checkers \& Fuzzers). 
}\label{tab:QUICstudyComparison}
\vspace{-4mm}
\begin{threeparttable}
\resizebox{\columnwidth}{!}{
\begin{tabular}{lccccccccccc} 
\toprule
\textbf{Tools}               & {\cellcolor[rgb]{0.882,0.882,0.882}}\begin{tabular}[c]{@{}>{\cellcolor[rgb]{0.882,0.882,0.882}}c@{}}\textbf{RFC}\\\textbf{9000/}\\\textbf{9001}\end{tabular} & \begin{tabular}[c]{@{}c@{}}\textbf{Open-}\\\textbf{source }\end{tabular} & \begin{tabular}[c]{@{}c@{}}\textbf{\textit{QUIC} Impls.}\\\textbf{tested }\end{tabular} & {\cellcolor[rgb]{0.882,0.882,0.882}}\textbf{\#Faults} & \begin{tabular}[c]{@{}c@{}}\textbf{Active}\\\textbf{learning }\end{tabular} & \begin{tabular}[c]{@{}c@{}}\textbf{Timing}\\\textbf{related}\\\textbf{behavior }\end{tabular} & \begin{tabular}[c]{@{}c@{}}\textbf{Invalid }\\\textbf{test }\\\textbf{cases }\end{tabular} & \begin{tabular}[c]{@{}c@{}}\textbf{Black }\\\textbf{box }\end{tabular} & \begin{tabular}[c]{@{}c@{}}\textbf{State }\\\textbf{model }\\\textbf{opt. }\end{tabular} & \begin{tabular}[c]{@{}c@{}}\textbf{Automated }\\\textbf{(w/o formal}\\\textbf{model)}\end{tabular} & \begin{tabular}[c]{@{}c@{}}\textbf{Under }\\\textbf{spec.}\\\textbf{ detection}\end{tabular}  \\ 
\midrule
\multicolumn{12}{l}{{\cellcolor[rgb]{1,0.925,0.792}}\textbf{Non-compliance checking}}                                                                                                                                                                                                                                                                                                                                                                                                                                                                                                                                                                                                                                                                                                                                                                                                                                                                                                                                                                                                 \\ 
\midrule
\textbf{\name{} (Ours)}                  & {\cellcolor[rgb]{0.882,0.882,0.882}}\ding{51}                                                                                                                       & \ding{51}                                                                       &                           \implementationcount{}                      & {\cellcolor[rgb]{0.882,0.882,0.882}}\bugcount{}      & \ding{51}                                                                          & \ding{51}                                                                                            & \ding{51}                                                                                         & \ding{51}                                                                     & \ding{51}                                                                                       & \ding{51}                                                                                                 & \ding{51}                                                                                            \\
EPIQ’2021\cite{crochet2021verifying}$\ddag$~~                  & {\cellcolor[rgb]{0.882,0.882,0.882}}\ding{55}                                                                                                                       & \ding{51}                                                                       & 7                                                                                       & {\cellcolor[rgb]{0.882,0.882,0.882}}9                 & \ding{55}                                                                          & \ding{55}                                                                                            & \ding{51}                                                                                         & \ding{51}                                                                    & -                                                                                        & \ding{55}                                                                                                 & \ding{51}                                                                                            \\
SIGCOMM'2021\cite{ferreira2021prognosis} & {\cellcolor[rgb]{0.882,0.882,0.882}}\ding{55} & \ding{51} & 3 & {\cellcolor[rgb]{0.882,0.882,0.882}}3 & \ding{51} & \ding{55} & \ding{55} & \ding{51} & \ding{55} & \ding{55} & \ding{51} \\
SIGCOMM’2019\cite{mcmillan2019formal}$\ddag$~~               & {\cellcolor[rgb]{0.882,0.882,0.882}}\ding{55}                                                                                                                       & \ding{51}                                                                       & 4                                                                                       & {\cellcolor[rgb]{0.882,0.882,0.882}}27                & \ding{55}                                                                          & \ding{55}                                                                                            & \ding{51}                                                                                         & \ding{51}                                                                     & -                                                                                        & \ding{55}                                                                                                 & \ding{51}                                                                                            \\
\textit{Unpublished}~(2019)~ & {\cellcolor[rgb]{0.882,0.882,0.882}}\ding{55}                                                                                                                       & \ding{51}                                                                       & 1                                                                                       & {\cellcolor[rgb]{0.882,0.882,0.882}}0                 & \ding{51}                                                                          & \ding{55}                                                                                            & \ding{55}                                                                                         & \ding{51}                                                                     & \ding{55}                                                                                       & \ding{55}                                                                                                 & \ding{55}                                                                                            \\
IMC’2017\cite{kakhki2017taking}\textasteriskcentered~~                   & {\cellcolor[rgb]{0.882,0.882,0.882}}\ding{55}                                                                                                                       & \ding{51}                                                                       & 1                                                                                       & {\cellcolor[rgb]{0.882,0.882,0.882}}1                 & \ding{55}                                                                          & \ding{55}                                                                                            & \ding{55}                                                                                         & \ding{55}                                                                     & \ding{55}                                                                                       & \ding{55}                                                                                                 & \ding{55}                                                                                            \\ 
\midrule
\multicolumn{12}{l}{{\cellcolor[rgb]{1,0.925,0.792}}\textbf{Other} testing studies (fuzzing, fault injection, traffic analysis)}                                                                                                                                                                                                                                                                                                                                                                                                                                                                                                                                                                                                                                                                                                                                                                                                                                                                                                                                                      \\ 
\midrule
QUIC-Fuzz~\cite{Quic-fuzz} (ESORICS'2025)$^\dagger$   & {\cellcolor[rgb]{0.882,0.882,0.882}}\ding{51}                                                                                                                       & \ding{51}                                                                       & 6                                                                                       & {\cellcolor[rgb]{0.882,0.882,0.882}}10                 & -                                                                           & \ding{55}                                                                                            & \ding{51}                                                                                         & \ding{55}                                                                     & -                                                                                        & -                                                                                                  & -                                                                                             \\
Fuzztruction-net~\cite{fuzztrution-net} (CCS'2024)$^\diamond$   & {\cellcolor[rgb]{0.882,0.882,0.882}}\ding{51}                                                                                                                       & \ding{51}                                                                       & 2                                                                                       & {\cellcolor[rgb]{0.882,0.882,0.882}}5                 & -                                                                           & \ding{51}                                                                                            & \ding{51}                                                                                         & \ding{55}                                                                     & -                                                                                        & -                                                                                                  & -                                                                                             \\
Bleem~\cite{luo2023bleem} (USENIX'2023)$^\dagger$               & {\cellcolor[rgb]{0.882,0.882,0.882}}\ding{51}                                                                                                                       & \ding{55}                                                                       & 1                                                                                      & {\cellcolor[rgb]{0.882,0.882,0.882}}1                 & -                                                                           & \ding{55}                                                                                            & \ding{51}                                                                                         & \ding{51}                                                                     & -                                                                                        & -                                                                                                  & -                                                                                             \\
DPIFuzz~\cite{reen2020dpifuzz} (ACSAC'2020)$^\star$             & {\cellcolor[rgb]{0.882,0.882,0.882}}\ding{55}                                                                                                                       & \ding{51}                                                                       & 5                                                                                       & {\cellcolor[rgb]{0.882,0.882,0.882}}4                 & -                                                                           & \ding{55}                                                                                            & \ding{51}                                                                                         & \ding{51}                                                                     & -                                                                                        & -                                                                                                  & \ding{51}                                                                                            \\
BooFuzz (\textit{Communit/Industry})$^\star$             & {\cellcolor[rgb]{0.882,0.882,0.882}}\ding{51}                                                                                                                       & \ding{51}                                                                       & -                                                                                       & {\cellcolor[rgb]{0.882,0.882,0.882}}-                 & -                                                                           & \ding{55}                                                                                            & \ding{51}                                                                                         & \ding{51}                                                                     & -                                                                                        & -                                                                                                  & -                                                                                             \\
\bottomrule
\end{tabular}
}
\begin{tablenotes}
\small
\item {\footnotesize$\ddag$:~Formal verification; $\diamond$: ~Fault-injection fuzzer; $\dagger$:~Mutation-based fuzzer.; $\star$~:~Generation-based fuzzer; \textasteriskcentered: Instruments a QUIC implementation to extract its execution trace.}
\end{tablenotes}
\end{threeparttable}
\end{table}
\end{minipage}
\end{figure*}

%%%%%%%%%%%%%%%%%%%%%%%%%%%%%%%%%%%%%%%%%%%%%%%%%%%%%%%%%%%%

%-------------------------------------------------------------------------------
\section{Related Work} \label{sec:relatedWork}
%-------------------------------------------------------------------------------
\noindent\textbf{Passive State Machine Inference.~}These approaches~\cite{passive-learning-app, passive-learning-app-2} employ template data, e.g. \texttt{pcap} files to generate an FSM. However, the approach cannot detect server failures (crashes) and actively explore \textit{unobserved} states of the system. Hence, we consider an active learning approach.

\noindent\textbf{Active State Machine Inference.~}Past studies used the approach to successfully test network protocols~\cite{fiteruau2016combining,RN17,RN51,RN21,RN22,mqttFuzzer, dtlsFuzzer, tlsFuzzer, hussain2021noncompliance}; a concurrent study employed the methods to successfully discover specification non-conformities of Bluetooth Low Energy~\cite{karim2023blediff} implementations. Notably, in~\cite{karim2023blediff}, a divide-and-conquer approach speeds up learning, which we do not consider for QUIC due to the possibility of the latter state transitions being influenced by the first inputs. A concurrent study also proposed an automated method~\cite{fiterau2023automata} to detect state machine bugs from learned models. The approach requires the construction of a Deterministic Finite Automaton first to describe the correct state transitions for the protocol handshake. In contrast, similar to the approach in \cite{fiteruau2016combining, RN17, mqttFuzzer, karim2023blediff}, our chosen method simplifies the analysis by directly performing cross-model checking to extract the deviations. Further, we also provide differential testing with curated reference models to reduce the burden of identifying non-compliant behaviors (see Figure~\ref{fig:QUIC-TeSter-use-case}).

\vspace{1px}
\noindent\textbf{Network Protocol Fuzzing.~}We acknowledge efforts in implementing mutation-based~\cite{aflnet, SGF, snapFuzz, maier2022fitm, nyxnet, luo2023bleem, fuzztrution-net,Quic-fuzz} and generation-based fuzzers~\cite{boofuzz, peach-fuzzer} for discovering memory-related bugs in protocols. A recent study~\cite{chatafl} leveraged large language models to enable structure-aware mutation on the non-security protocol specifications (RFCs). In contrast, our work focuses on employing automata learning as a black-box tool for uncovering specification violations in protocol implementations.

\vspace{-1mm}
%-------------------------------------------------------------------------------
\section{Discussion} \label{sec:discussion}
%-------------------------------------------------------------------------------
\noindent\textbf{Automated Compliance Testing With Reference FSMs.~}Through our differential analysis method (Temporal and Cross Model, illustrated in Figure~\ref{fig:differentialTestingOnTemporal}), we identified and curated \referenceModelCount{} models conforming to QUIC RFC 9000/9001 (see Appendix~\ref{sec:refmode-appendix}). The reference models capture compliant FSM variations across all 5 security configurations. Importantly, the models allow developers to employ \name{} to effectively, efficiently and automatically identify non-conformance behaviors or faults in the future as illustrated in Figure~\ref{fig:QUIC-TeSter-use-case}.

\vspace{2px}
\noindent\textbf{Threats to validity.~}
We uncovered \bugcount{} faults and confirmed all of the deviations detected by \differentialComparator{} are due to either: i)~specification bugs, ii)~memory-corruption bugs, or iii)~logical flaws as listed in Table~\ref{tab:bug-table-summary}. \name{} did not detect any false positives; identifying a deviation that does not lead to a fault. Further, we have made efforts to ensure the series of membership queries produced based on our dictionary of symbols comprehensively covers all protocol variations. However, a non-zero probability exists, despite our best efforts, that we have inadvertently missed a symbolization and hence, a potential uncovering of a state or transition. Further, it is important to recognize the test oracles (diverse QUIC implementations) are inherently unfaithful because they can all suffer from the same logical vulnerability. Thus, although highly unlikely, it is possible that a noncompliance remains, yet undiscovered despite the diverse range of implementations and settings we employed in our testing.

Notably, the noncompliance checker is not purposed for detecting memory-corruption bugs effectively because active automata learning focuses on the logical structure and behavior of the state machine rather than the underlying data manipulation. For instance, it does not perform mutations on packet fields that could trigger buffer overflows or similar vulnerabilities. 

\vspace{2px}
\noindent\textbf{Correctness.~}
As highlighted in Section~\ref{sec:experiments}, we have manually validated all 186 models and confirmed that all the behaviors illustrated in the models (across 19 implementations) are reproducible. In addition, we provide a list of inputs for each deviating behavior in our code repository for reproducibility~\cite {repo-inputs}. As an interesting anecdote, we spent approximately 1860 mins (approximately 10 mins per model for 186 models) in our efforts to evaluate the veracity of the state machine comparison method discussed in Section~\ref{sec:differentialTesting}.

We also considered if the optimizations are sound and produce FSMs that accept the same strings as the un-optimized FSMs. To this end, we applied the differential analysis method on un-optimize FSMs, for the set of models we evaluated, the task consumes approximately 5.4 hours. Subsequently, we compared the unique deviations extracted with those deviations extracted from optimized FSMs (taking approximately 2.2 hours). We found the unique deviations from the un-optimised and optimised models to be the same.

\vspace{2px}
\noindent\textbf{Completeness.~}In this study, we focused on comprehensively analysing the security components of QUIC (the handshake stages). Hence, \name{} is not currently capable of detecting deviations in other components of QUIC. These include connection migration of a client on an active connection with a server or the state of open streams for application data exchange. But, \name{} is modular and extensible to test these components, it will require defining new inputs and output symbols and modifying the existing \mapper{}. We leave these avenues for further development.

\noindent\textbf{Learning Time-Dependent Behaviour.} As explained in Section~\ref{sec:picoquic_memory}, learning with time-parametrised inputs can uncover previously unobserved behaviours, lead to the discovery of new states, state transitions, and potential new bug discoveries such as \textbf{M}-4. However, the average time required for learning increases with different time settings---29.6 hours for $short$, 38.9 hours for $long$, and 76.5 hours for both $short$ and $long$. Therefore, exploring optimization strategies to reduce runtime while maintaining the bug discovery effectiveness is a valuable avenue for future work.

\vspace{2px}
\noindent\textbf{Comparison with existing tools.~}Prior to the QUIC specifications~\cite{9000,9001} being finalized, \cite{RN15, crochet2021verifying, kakhki2017taking, mcmillan2019formal, reen2020dpifuzz,ferreira2021prognosis} have made significant efforts in testing QUIC implementations. However, most of the QUIC-specific tools are only built for testing the IETF-draft or Google-QUIC, are no longer actively maintained, and have limited scope for security testing (e.g. do not consider invalid input while testing). Therefore, these tools lack suitability for testing the ratified specification. To the best of our knowledge, our work is \textit{first} to develop a black-box noncompliance checker to test \textit{all 5 different security settings}, including client address validation and client authentication of the ratified QUIC specification. We summarize our comparison in Table~\ref{tab:QUICstudyComparison} and, \textit{for completeness}, include mutation-based fuzzers and a generation-based fuzzers that have tested QUIC implementations. 
%-------------------------------------------------------------------------------
%\vspace{-3mm}
\section{Conclusions and Future Work} \label{sec:conclusion}
%----------------------------------------------------------------------
In this study, we presented the first, programming language agnostic, comprehensive noncompliance tester for the security critical connection establishment components of the QUIC protocol. \name{} is validated with 5 CVEs assigned from developers, \bugcount{} faults discovered and confirmed by developers, a bug bounty and uncovering of a specification ambiguity. Although we have made significant in-roads (our \opt{} and \differentialComparator{}) to reduce the manual effort required to analyze the generated models, exploring automatic model analysis techniques leveraging LLMs could further assist in model analysis to reduce dependence on domain expertise. Further, extending \name{} to evaluate non-security components of QUIC are avenues for future work.

\bibliographystyle{ACM-Reference-Format}
\bibliography{references}


\begin{thebibliography}{65}

%%% the Web address.

\ifx \showCODEN    \undefined \def \showCODEN     #1{\unskip}     \fi
\ifx \showDOI      \undefined \def \showDOI       #1{#1}\fi
\ifx \showISBNx    \undefined \def \showISBNx     #1{\unskip}     \fi
\ifx \showISBNxiii \undefined \def \showISBNxiii  #1{\unskip}     \fi
\ifx \showISSN     \undefined \def \showISSN      #1{\unskip}     \fi
\ifx \showLCCN     \undefined \def \showLCCN      #1{\unskip}     \fi
\ifx \shownote     \undefined \def \shownote      #1{#1}          \fi
\ifx \showarticletitle \undefined \def \showarticletitle #1{#1}   \fi
\ifx \showURL      \undefined \def \showURL       {\relax}        \fi
% invisible to TeX
\providecommand\bibfield[2]{#2}
\providecommand\bibinfo[2]{#2}
\providecommand\natexlab[1]{#1}
\providecommand\showeprint[2][]{arXiv:#2}

\bibitem[aio({[n.\,d.]})]%
        {aioquic}
 \bibinfo{year}{[n.\,d.]}\natexlab{}.
\newblock \bibinfo{title}{Aioquic}.
\newblock \bibinfo{howpublished}{\url{https://aioquic.readthedocs.io/en/latest/}}.
\newblock
\newblock
\shownote{Accessed: 10 October 2022}.


\bibitem[bor({[n.\,d.]})]%
        {boringssl}
 \bibinfo{year}{[n.\,d.]}\natexlab{}.
\newblock \bibinfo{title}{BoringSSL}.
\newblock \bibinfo{howpublished}{\url{https://boringssl.googlesource.com/boringssl/}}.
\newblock
\newblock
\shownote{Accessed: 7 June 2022}.


\bibitem[rep({[n.\,d.]})]%
        {repo-inputs}
 \bibinfo{year}{[n.\,d.]}\natexlab{}.
\newblock \bibinfo{title}{Bug Description with input sequence to reproduce the faults}.
\newblock \bibinfo{howpublished}{\url{https://anonymous.4open.science/r/QUICTester-7EBC/results/README.md}}.
\newblock
\newblock
\shownote{Accessed: 2 August 2024}.


\bibitem[rr({[n.\,d.]})]%
        {rr}
 \bibinfo{year}{[n.\,d.]}\natexlab{}.
\newblock \bibinfo{title}{rr: lightweight recording \& deterministic debugging}.
\newblock \bibinfo{howpublished}{\url{https://rr-project.org/}}.
\newblock
\newblock
\shownote{Accessed: 16 January 2023}.


\bibitem[QUI({[n.\,d.]})]%
        {QUICstats}
 \bibinfo{year}{[n.\,d.]}\natexlab{}.
\newblock \bibinfo{title}{Usage statistics of HTTP/3 for websites}.
\newblock \bibinfo{howpublished}{\url{https://w3techs.com/technologies/details/ce-http3}}.
\newblock
\newblock
\shownote{Accessed: 7 June 2023}.


\bibitem[Aichernig et~al\mbox{.}(2021)]%
        {mqttFuzzer}
\bibfield{author}{\bibinfo{person}{Bernhard~K Aichernig}, \bibinfo{person}{Edi Muškardin}, {and} \bibinfo{person}{Andrea Pferscher}.} \bibinfo{year}{2021}\natexlab{}.
\newblock \showarticletitle{Learning-based fuzzing of {IoT} message brokers}. In \bibinfo{booktitle}{\emph{IEEE Conference on Software Testing, Verification and Validation (ICST)}}. \bibinfo{pages}{47--58}.
\newblock
\showISBNx{1728168368}


\bibitem[Andronidis and Cadar(2022)]%
        {snapFuzz}
\bibfield{author}{\bibinfo{person}{Anastasios Andronidis} {and} \bibinfo{person}{Cristian Cadar}.} \bibinfo{year}{2022}\natexlab{}.
\newblock \showarticletitle{{SnapFuzz}: high-throughput fuzzing of network applications}. In \bibinfo{booktitle}{\emph{ACM SIGSOFT International Symposium on Software Testing and Analysis (ISSTA)}}. \bibinfo{pages}{340--351}.
\newblock


\bibitem[Ang and Ranasinghe(2025)]%
        {Quic-fuzz}
\bibfield{author}{\bibinfo{person}{Kian~Kai Ang} {and} \bibinfo{person}{Damith~C. Ranasinghe}.} \bibinfo{year}{2025}\natexlab{}.
\newblock \showarticletitle{QUIC-Fuzz: An Effective Greybox Fuzzer For The QUIC Protocol}. In \bibinfo{booktitle}{\emph{European Symposium on Research in Computer Security (ESORICS)}}.
\newblock


\bibitem[Angluin(1987)]%
        {angluin1987learning}
\bibfield{author}{\bibinfo{person}{Dana Angluin}.} \bibinfo{year}{1987}\natexlab{}.
\newblock \showarticletitle{Learning regular sets from queries and counterexamples}.
\newblock \bibinfo{journal}{\emph{Information and computation}} \bibinfo{volume}{75}, \bibinfo{number}{2} (\bibinfo{year}{1987}), \bibinfo{pages}{87--106}.
\newblock


\bibitem[Ba et~al\mbox{.}(2022)]%
        {SGF}
\bibfield{author}{\bibinfo{person}{Jinsheng Ba}, \bibinfo{person}{Marcel B{\"o}hme}, \bibinfo{person}{Zahra Mirzamomen}, {and} \bibinfo{person}{Abhik Roychoudhury}.} \bibinfo{year}{2022}\natexlab{}.
\newblock \showarticletitle{Stateful greybox fuzzing}. In \bibinfo{booktitle}{\emph{USENIX Security Symposium (USENIX Security)}}. \bibinfo{pages}{3255--3272}.
\newblock


\bibitem[Bars et~al\mbox{.}(2024)]%
        {fuzztrution-net}
\bibfield{author}{\bibinfo{person}{Nils Bars}, \bibinfo{person}{Moritz Schloegel}, \bibinfo{person}{Nico Schiller}, \bibinfo{person}{Lukas Bernhard}, {and} \bibinfo{person}{Thorsten Holz}.} \bibinfo{year}{2024}\natexlab{}.
\newblock \showarticletitle{No Peer, no Cry: Network Application Fuzzing via Fault Injection}.
\newblock  (\bibinfo{year}{2024}).
\newblock


\bibitem[Beurdouche et~al\mbox{.}(2015)]%
        {7163046}
\bibfield{author}{\bibinfo{person}{Benjamin Beurdouche}, \bibinfo{person}{Karthikeyan Bhargavan}, \bibinfo{person}{Antoine Delignat-Lavaud}, \bibinfo{person}{Cédric Fournet}, \bibinfo{person}{Markulf Kohlweiss}, \bibinfo{person}{Alfredo Pironti}, \bibinfo{person}{Pierre-Yves Strub}, {and} \bibinfo{person}{Jean~Karim Zinzindohoue}.} \bibinfo{year}{2015}\natexlab{}.
\newblock \showarticletitle{A Messy State of the Union: Taming the Composite State Machines of TLS}. In \bibinfo{booktitle}{\emph{IEEE Symposium on Security and Privacy (S\&P)}}. \bibinfo{pages}{535--552}.
\newblock
\urldef\tempurl%
\url{https://doi.org/10.1109/SP.2015.39}
\showDOI{\tempurl}


\bibitem[Brown et~al\mbox{.}(2020)]%
        {in-context-few-short-learning-1}
\bibfield{author}{\bibinfo{person}{Tom Brown}, \bibinfo{person}{Benjamin Mann}, \bibinfo{person}{Nick Ryder}, \bibinfo{person}{Melanie Subbiah}, \bibinfo{person}{Jared~D Kaplan}, \bibinfo{person}{Prafulla Dhariwal}, \bibinfo{person}{Arvind Neelakantan}, \bibinfo{person}{Pranav Shyam}, \bibinfo{person}{Girish Sastry}, \bibinfo{person}{Amanda Askell}, {et~al\mbox{.}}} \bibinfo{year}{2020}\natexlab{}.
\newblock \showarticletitle{Language models are few-shot learners}.
\newblock \bibinfo{journal}{\emph{Advances in neural information processing systems}}  \bibinfo{volume}{33} (\bibinfo{year}{2020}), \bibinfo{pages}{1877--1901}.
\newblock


\bibitem[Brubaker et~al\mbox{.}(2014)]%
        {6956560}
\bibfield{author}{\bibinfo{person}{Chad Brubaker}, \bibinfo{person}{Suman Jana}, \bibinfo{person}{Baishakhi Ray}, \bibinfo{person}{Sarfraz Khurshid}, {and} \bibinfo{person}{Vitaly Shmatikov}.} \bibinfo{year}{2014}\natexlab{}.
\newblock \showarticletitle{Using Frankencerts for Automated Adversarial Testing of Certificate Validation in SSL/TLS Implementations}. In \bibinfo{booktitle}{\emph{IEEE Symposium on Security and Privacy (S\&P)}}. \bibinfo{pages}{114--129}.
\newblock
\urldef\tempurl%
\url{https://doi.org/10.1109/SP.2014.15}
\showDOI{\tempurl}


\bibitem[Chen et~al\mbox{.}(2023)]%
        {chen2023can}
\bibfield{author}{\bibinfo{person}{Yufan Chen}, \bibinfo{person}{Arjun Arunasalam}, {and} \bibinfo{person}{Z~Berkay Celik}.} \bibinfo{year}{2023}\natexlab{}.
\newblock \showarticletitle{Can large language models provide security \& privacy advice? measuring the ability of llms to refute misconceptions}. In \bibinfo{booktitle}{\emph{Annual Computer Security Applications Conference (ACSAC)}}. \bibinfo{pages}{366--378}.
\newblock


\bibitem[Chow(1978)]%
        {wpmethod}
\bibfield{author}{\bibinfo{person}{Tsun~S. Chow}.} \bibinfo{year}{1978}\natexlab{}.
\newblock \showarticletitle{Testing software design modeled by finite-state machines}.
\newblock \bibinfo{journal}{\emph{IEEE transactions on software engineering}} \bibinfo{number}{3} (\bibinfo{year}{1978}), \bibinfo{pages}{178--187}.
\newblock


\bibitem[Comparetti et~al\mbox{.}(2009)]%
        {passive-learning-app}
\bibfield{author}{\bibinfo{person}{Paolo~Milani Comparetti}, \bibinfo{person}{Gilbert Wondracek}, \bibinfo{person}{Christopher Kruegel}, {and} \bibinfo{person}{Engin Kirda}.} \bibinfo{year}{2009}\natexlab{}.
\newblock \showarticletitle{Prospex: Protocol specification extraction}. In \bibinfo{booktitle}{\emph{IEEE Symposium on Security and Privacy (S\&P)}}. IEEE, \bibinfo{pages}{110--125}.
\newblock


\bibitem[Crochet et~al\mbox{.}(2021)]%
        {crochet2021verifying}
\bibfield{author}{\bibinfo{person}{Christophe Crochet}, \bibinfo{person}{Tom Rousseaux}, \bibinfo{person}{Maxime Piraux}, \bibinfo{person}{Jean-Fran{\c{c}}ois Sambon}, {and} \bibinfo{person}{Axel Legay}.} \bibinfo{year}{2021}\natexlab{}.
\newblock \showarticletitle{Verifying QUIC implementations using Ivy}. In \bibinfo{booktitle}{\emph{Proceedings of the 2021 Workshop on Evolution, Performance and Interoperability of QUIC}}. \bibinfo{pages}{35--41}.
\newblock


\bibitem[Daniel et~al\mbox{.}(2018)]%
        {RN21}
\bibfield{author}{\bibinfo{person}{Lesly-Ann Daniel}, \bibinfo{person}{Erik Poll}, {and} \bibinfo{person}{Joeri de Ruiter}.} \bibinfo{year}{2018}\natexlab{}.
\newblock \showarticletitle{Inferring {OpenVPN} state machines using protocol state fuzzing}. In \bibinfo{booktitle}{\emph{IEEE European Symposium On Security And Privacy Workshops (EuroS\&PW)}}. \bibinfo{pages}{11--19}.
\newblock
\showISBNx{1538654458}


\bibitem[De~Ruiter and Poll(2015)]%
        {tlsFuzzer}
\bibfield{author}{\bibinfo{person}{Joeri De~Ruiter} {and} \bibinfo{person}{Erik Poll}.} \bibinfo{year}{2015}\natexlab{}.
\newblock \showarticletitle{Protocol state fuzzing of {TLS} implementations}. In \bibinfo{booktitle}{\emph{USENIX Security Symposium (USENIX Security)}}. \bibinfo{pages}{193--206}.
\newblock
\showISBNx{1939133114}


\bibitem[Deng et~al\mbox{.}(2023)]%
        {llm-zero-shot-deep-lib}
\bibfield{author}{\bibinfo{person}{Yinlin Deng}, \bibinfo{person}{Chunqiu~Steven Xia}, \bibinfo{person}{Haoran Peng}, \bibinfo{person}{Chenyuan Yang}, {and} \bibinfo{person}{Lingming Zhang}.} \bibinfo{year}{2023}\natexlab{}.
\newblock \showarticletitle{Large language models are zero-shot fuzzers: Fuzzing deep-learning libraries via large language models}. In \bibinfo{booktitle}{\emph{ACM SIGSOFT International Symposium on Software Testing and Analysis (ISSTA)}}. \bibinfo{pages}{423--435}.
\newblock


\bibitem[Dierks and Rescorla(2008a)]%
        {RN49}
\bibfield{author}{\bibinfo{person}{Tim Dierks} {and} \bibinfo{person}{Eric Rescorla}.} \bibinfo{year}{2008}\natexlab{a}.
\newblock \bibinfo{title}{The transport layer security (TLS) protocol version 1.2}.
\newblock \bibinfo{howpublished}{RFC 5246}.
\newblock


\bibitem[Dierks and Rescorla(2008b)]%
        {TLS1.2RFC}
\bibfield{author}{\bibinfo{person}{Tim Dierks} {and} \bibinfo{person}{Eric Rescorla}.} \bibinfo{year}{2008}\natexlab{b}.
\newblock \bibinfo{title}{The transport layer security (TLS) protocol version 1.2}.
\newblock \bibinfo{howpublished}{RFC 5246}.
\newblock


\bibitem[Eddington({[n.\,d.]})]%
        {peach-fuzzer}
\bibfield{author}{\bibinfo{person}{Michael Eddington}.} \bibinfo{year}{[n.\,d.]}\natexlab{}.
\newblock \bibinfo{title}{Peach fuzzing platform}.
\newblock \bibinfo{howpublished}{\url{https://gitlab.com/gitlab-org/security-products/protocol-fuzzer-ce.}}.
\newblock
\newblock
\shownote{Accessed: 2 August 2024}.


\bibitem[Fern{\'a}ndez et~al\mbox{.}(2020)]%
        {iotQuicProve}
\bibfield{author}{\bibinfo{person}{F{\'a}tima Fern{\'a}ndez}, \bibinfo{person}{Mihail Zverev}, \bibinfo{person}{Pablo Garrido}, \bibinfo{person}{Jos{\'e}~R Ju{\'a}rez}, \bibinfo{person}{Josu Bilbao}, {and} \bibinfo{person}{Ram{\'o}n Ag{\"u}ero}.} \bibinfo{year}{2020}\natexlab{}.
\newblock \showarticletitle{{And QUIC meets IoT: performance assessment of MQTT over QUIC}}. In \bibinfo{booktitle}{\emph{International Conference on Wireless and Mobile Computing, Networking and Communications (WiMob)}}.
\newblock


\bibitem[Ferreira et~al\mbox{.}(2021)]%
        {ferreira2021prognosis}
\bibfield{author}{\bibinfo{person}{Tiago Ferreira}, \bibinfo{person}{Harrison Brewton}, \bibinfo{person}{Loris D'Antoni}, {and} \bibinfo{person}{Alexandra Silva}.} \bibinfo{year}{2021}\natexlab{}.
\newblock \showarticletitle{Prognosis: closed-box analysis of network protocol implementations}. In \bibinfo{booktitle}{\emph{Proceedings of the 2021 ACM SIGCOMM 2021 Conference}}. \bibinfo{pages}{762--774}.
\newblock


\bibitem[Fiter{\u{a}}u-Bro{\c{s}}tean et~al\mbox{.}(2016)]%
        {fiteruau2016combining}
\bibfield{author}{\bibinfo{person}{Paul Fiter{\u{a}}u-Bro{\c{s}}tean}, \bibinfo{person}{Ramon Janssen}, {and} \bibinfo{person}{Frits Vaandrager}.} \bibinfo{year}{2016}\natexlab{}.
\newblock \showarticletitle{Combining model learning and model checking to analyze TCP implementations}. In \bibinfo{booktitle}{\emph{International Conference on Computer Aided Verification (CAV)}}.
\newblock


\bibitem[Fiterau-Brostean et~al\mbox{.}(2020)]%
        {dtlsFuzzer}
\bibfield{author}{\bibinfo{person}{Paul Fiterau-Brostean}, \bibinfo{person}{Bengt Jonsson}, \bibinfo{person}{Robert Merget}, \bibinfo{person}{Joeri De~Ruiter}, \bibinfo{person}{Konstantinos Sagonas}, {and} \bibinfo{person}{Juraj Somorovsky}.} \bibinfo{year}{2020}\natexlab{}.
\newblock \showarticletitle{Analysis of {DTLS} implementations using protocol state fuzzing}. In \bibinfo{booktitle}{\emph{USENIX Security Symposium (USENIX Security)}}. \bibinfo{pages}{2523--2540}.
\newblock
\showISBNx{1939133173}


\bibitem[Fiterau-Brostean et~al\mbox{.}(2023)]%
        {fiterau2023automata}
\bibfield{author}{\bibinfo{person}{Paul Fiterau-Brostean}, \bibinfo{person}{Bengt Jonsson}, \bibinfo{person}{Konstantinos Sagonas}, {and} \bibinfo{person}{Fredrik T{\aa}quist}.} \bibinfo{year}{2023}\natexlab{}.
\newblock \showarticletitle{Automata-Based Automated Detection of State Machine Bugs in Protocol Implementations.}. In \bibinfo{booktitle}{\emph{Network and Distributed System Security (NDSS)}}.
\newblock


\bibitem[Fiterău-Broştean et~al\mbox{.}(2017)]%
        {RN17}
\bibfield{author}{\bibinfo{person}{Paul Fiterău-Broştean}, \bibinfo{person}{Toon Lenaerts}, \bibinfo{person}{Erik Poll}, \bibinfo{person}{Joeri de Ruiter}, \bibinfo{person}{Frits Vaandrager}, {and} \bibinfo{person}{Patrick Verleg}.} \bibinfo{year}{2017}\natexlab{}.
\newblock \showarticletitle{Model learning and model checking of SSH implementations}. In \bibinfo{booktitle}{\emph{ACM SIGSOFT International Symposium on Model Checking of Software (SPIN)}}. \bibinfo{pages}{142--151}.
\newblock


\bibitem[Grinchtein et~al\mbox{.}(2010)]%
        {grinchtein2010learning}
\bibfield{author}{\bibinfo{person}{Olga Grinchtein}, \bibinfo{person}{Bengt Jonsson}, {and} \bibinfo{person}{Martin Leucker}.} \bibinfo{year}{2010}\natexlab{}.
\newblock \showarticletitle{Learning of event-recording automata}.
\newblock \bibinfo{journal}{\emph{Theoretical Computer Science}} \bibinfo{volume}{411}, \bibinfo{number}{47} (\bibinfo{year}{2010}), \bibinfo{pages}{4029--4054}.
\newblock


\bibitem[Hou et~al\mbox{.}(2022)]%
        {hou2022qfaas}
\bibfield{author}{\bibinfo{person}{Kaiyu Hou}, \bibinfo{person}{Sen Lin}, \bibinfo{person}{Yan Chen}, {and} \bibinfo{person}{Vinod Yegneswaran}.} \bibinfo{year}{2022}\natexlab{}.
\newblock \showarticletitle{{QFaaS}: accelerating and securing serverless cloud networks with {QUIC}}. In \bibinfo{booktitle}{\emph{Symposium on Cloud Computing (SoCC)}}. \bibinfo{pages}{240--256}.
\newblock


\bibitem[Hsu et~al\mbox{.}(2008)]%
        {passive-learning-app-2}
\bibfield{author}{\bibinfo{person}{Yating Hsu}, \bibinfo{person}{Guoqiang Shu}, {and} \bibinfo{person}{David Lee}.} \bibinfo{year}{2008}\natexlab{}.
\newblock \showarticletitle{A model-based approach to security flaw detection of network protocol implementations}. In \bibinfo{booktitle}{\emph{IEEE International Conference on Network Protocols}}. IEEE, \bibinfo{pages}{114--123}.
\newblock


\bibitem[Huitema et~al\mbox{.}(2022)]%
        {9250}
\bibfield{author}{\bibinfo{person}{Christian Huitema}, \bibinfo{person}{Sara Dickinson}, {and} \bibinfo{person}{Allison Mankin}.} \bibinfo{year}{2022}\natexlab{}.
\newblock \bibinfo{title}{{DNS over Dedicated QUIC Connections}}.
\newblock \bibinfo{howpublished}{RFC 9250}.
\newblock
\urldef\tempurl%
\url{https://doi.org/10.17487/RFC9250}
\showDOI{\tempurl}


\bibitem[Hussain et~al\mbox{.}(2021)]%
        {hussain2021noncompliance}
\bibfield{author}{\bibinfo{person}{Syed~Rafiul Hussain}, \bibinfo{person}{Imtiaz Karim}, \bibinfo{person}{Abdullah~Al Ishtiaq}, \bibinfo{person}{Omar Chowdhury}, {and} \bibinfo{person}{Elisa Bertino}.} \bibinfo{year}{2021}\natexlab{}.
\newblock \showarticletitle{Noncompliance as deviant behavior: An automated black-box noncompliance checker for {4G} {LTE} cellular devices}. In \bibinfo{booktitle}{\emph{Proceedings of the 2021 ACM SIGSAC Conference on Computer and Communications Security}}. \bibinfo{pages}{1082--1099}.
\newblock


\bibitem[{IoT Analytics GmbH}(2023)]%
        {state_of_iot}
\bibfield{author}{\bibinfo{person}{{IoT Analytics GmbH}}.} \bibinfo{year}{2023}\natexlab{}.
\newblock \bibinfo{title}{State of {IoT} 2023: Number of connected {IoT} devices growing 16\% to 16.7 billion globally}.
\newblock \bibinfo{howpublished}{\url{https://iot-analytics.com/number-connected-iot-devices}}.
\newblock


\bibitem[Isberner(2015)]%
        {isberner2015foundations}
\bibfield{author}{\bibinfo{person}{Malte Isberner}.} \bibinfo{year}{2015}\natexlab{}.
\newblock \showarticletitle{Foundations of active automata learning: an algorithmic perspective}.
\newblock  (\bibinfo{year}{2015}).
\newblock


\bibitem[Isberner et~al\mbox{.}(2014)]%
        {isberner2014ttt}
\bibfield{author}{\bibinfo{person}{Malte Isberner}, \bibinfo{person}{Falk Howar}, {and} \bibinfo{person}{Bernhard Steffen}.} \bibinfo{year}{2014}\natexlab{}.
\newblock \showarticletitle{The TTT algorithm: a redundancy-free approach to active automata learning}. In \bibinfo{booktitle}{\emph{Runtime Verification: 5th International Conference, RV 2014, Toronto, ON, Canada, September 22-25, 2014. Proceedings 5}}. Springer, \bibinfo{pages}{307--322}.
\newblock


\bibitem[Isberner et~al\mbox{.}(2015)]%
        {learnlib}
\bibfield{author}{\bibinfo{person}{Malte Isberner}, \bibinfo{person}{Falk Howar}, {and} \bibinfo{person}{Bernhard Steffen}.} \bibinfo{year}{2015}\natexlab{}.
\newblock \showarticletitle{The open-source learnLib: a framework for active automata learning}. In \bibinfo{booktitle}{\emph{Computer Aided Verification: 27th International Conference, CAV 2015, San Francisco, CA, USA, July 18-24, 2015, Proceedings, Part I 27}}. Springer, \bibinfo{pages}{487--495}.
\newblock


\bibitem[Iyengar and Thomson(2021)]%
        {9000}
\bibfield{author}{\bibinfo{person}{J Iyengar} {and} \bibinfo{person}{M Thomson}.} \bibinfo{year}{2021}\natexlab{}.
\newblock \bibinfo{title}{{QUIC: A UDP-Based Multiplexed and Secure Transport}}.
\newblock \bibinfo{howpublished}{RFC 9000}.
\newblock


\bibitem[jtpereyda({[n.\,d.]})]%
        {boofuzz}
\bibfield{author}{\bibinfo{person}{jtpereyda}.} \bibinfo{year}{[n.\,d.]}\natexlab{}.
\newblock \showarticletitle{BooFuzz: Network protocol fuzzing for humans}. \bibinfo{howpublished}{https://github.com/jtpereyda/boofuzz}.
\newblock


\bibitem[Kakhki et~al\mbox{.}(2017)]%
        {kakhki2017taking}
\bibfield{author}{\bibinfo{person}{Arash~Molavi Kakhki}, \bibinfo{person}{Samuel Jero}, \bibinfo{person}{David Choffnes}, \bibinfo{person}{Cristina Nita-Rotaru}, {and} \bibinfo{person}{Alan Mislove}.} \bibinfo{year}{2017}\natexlab{}.
\newblock \showarticletitle{Taking a long look at QUIC: an approach for rigorous evaluation of rapidly evolving transport protocols}. In \bibinfo{booktitle}{\emph{Proceedings of the Internet Measurement Conference}}. \bibinfo{pages}{290--303}.
\newblock


\bibitem[Kang et~al\mbox{.}(2023)]%
        {llvm-few-short-bug-repro}
\bibfield{author}{\bibinfo{person}{Sungmin Kang}, \bibinfo{person}{Juyeon Yoon}, {and} \bibinfo{person}{Shin Yoo}.} \bibinfo{year}{2023}\natexlab{}.
\newblock \showarticletitle{Large language models are few-shot testers: Exploring llm-based general bug reproduction}. In \bibinfo{booktitle}{\emph{2023 IEEE/ACM 45th International Conference on Software Engineering (ICSE)}}. IEEE, \bibinfo{pages}{2312--2323}.
\newblock


\bibitem[Karim et~al\mbox{.}(2023)]%
        {karim2023blediff}
\bibfield{author}{\bibinfo{person}{Imtiaz Karim}, \bibinfo{person}{Abdullah Al~Ishtiaq}, \bibinfo{person}{Syed~Rafiul Hussain}, {and} \bibinfo{person}{Elisa Bertino}.} \bibinfo{year}{2023}\natexlab{}.
\newblock \showarticletitle{BLEDiff: Scalable and Property-Agnostic Noncompliance Checking for BLE Implementations}. In \bibinfo{booktitle}{\emph{IEEE Symposium on Security and Privacy (S\&P)}}. IEEE, \bibinfo{pages}{3209--3227}.
\newblock


\bibitem[Kumar and Dezfouli(2019)]%
        {kumar2019implementation}
\bibfield{author}{\bibinfo{person}{Puneet Kumar} {and} \bibinfo{person}{Behnam Dezfouli}.} \bibinfo{year}{2019}\natexlab{}.
\newblock \showarticletitle{{Implementation and analysis of QUIC for MQTT}}.
\newblock \bibinfo{journal}{\emph{Computer Networks}}  \bibinfo{volume}{150} (\bibinfo{year}{2019}), \bibinfo{pages}{28--45}.
\newblock


\bibitem[Luo et~al\mbox{.}(2023)]%
        {luo2023bleem}
\bibfield{author}{\bibinfo{person}{Zhengxiong Luo}, \bibinfo{person}{Junze Yu}, \bibinfo{person}{Feilong Zuo}, \bibinfo{person}{Jianzhong Liu}, \bibinfo{person}{Yu Jiang}, \bibinfo{person}{Ting Chen}, \bibinfo{person}{Abhik Roychoudhury}, {and} \bibinfo{person}{Jiaguang Sun}.} \bibinfo{year}{2023}\natexlab{}.
\newblock \showarticletitle{Bleem: Packet sequence oriented fuzzing for protocol implementations}. In \bibinfo{booktitle}{\emph{USENIX Security Symposium (USENIX Security)}}. \bibinfo{pages}{4481--4498}.
\newblock


\bibitem[Maier et~al\mbox{.}(2022)]%
        {maier2022fitm}
\bibfield{author}{\bibinfo{person}{Dominik Maier}, \bibinfo{person}{Otto Bittner}, \bibinfo{person}{Marc Munier}, {and} \bibinfo{person}{Julian Beier}.} \bibinfo{year}{2022}\natexlab{}.
\newblock \showarticletitle{{FitM}: Binary-Only Coverage-Guided Fuzzing for Stateful Network Protocols}. In \bibinfo{booktitle}{\emph{Workshop on Binary Analysis Research (BAR)}}.
\newblock


\bibitem[McMahon~Stone et~al\mbox{.}(2018)]%
        {RN22}
\bibfield{author}{\bibinfo{person}{Chris McMahon~Stone}, \bibinfo{person}{Tom Chothia}, {and} \bibinfo{person}{Joeri~de Ruiter}.} \bibinfo{year}{2018}\natexlab{}.
\newblock \showarticletitle{Extending automated protocol state learning for the 802.11 4-way handshake}. In \bibinfo{booktitle}{\emph{European Symposium on Research in Computer Security (ESORICS)}}. \bibinfo{pages}{325--345}.
\newblock


\bibitem[McMillan and Zuck(2019)]%
        {mcmillan2019formal}
\bibfield{author}{\bibinfo{person}{Kenneth~L McMillan} {and} \bibinfo{person}{Lenore~D Zuck}.} \bibinfo{year}{2019}\natexlab{}.
\newblock \showarticletitle{Formal specification and testing of QUIC}.
\newblock In \bibinfo{booktitle}{\emph{Proceedings of the ACM Special Interest Group on Data Communication}}. \bibinfo{pages}{227--240}.
\newblock


\bibitem[Meng et~al\mbox{.}(2024)]%
        {chatafl}
\bibfield{author}{\bibinfo{person}{Ruijie Meng}, \bibinfo{person}{Martin Mirchev}, \bibinfo{person}{Marcel B{\"o}hme}, {and} \bibinfo{person}{Abhik Roychoudhury}.} \bibinfo{year}{2024}\natexlab{}.
\newblock \showarticletitle{Large language model guided protocol fuzzing}. In \bibinfo{booktitle}{\emph{Network and Distributed System Security (NDSS)}}.
\newblock


\bibitem[Pham et~al\mbox{.}(2020)]%
        {aflnet}
\bibfield{author}{\bibinfo{person}{Van-Thuan Pham}, \bibinfo{person}{Marcel Böhme}, {and} \bibinfo{person}{Abhik Roychoudhury}.} \bibinfo{year}{2020}\natexlab{}.
\newblock \showarticletitle{AFLNet: a greybox fuzzer for network protocols}. In \bibinfo{booktitle}{\emph{IEEE International Conference on Software Testing, Validation and Verification (ICST)}}. \bibinfo{pages}{460--465}.
\newblock
\showISBNx{1728157781}


\bibitem[Postel(1981)]%
        {RN43}
\bibfield{author}{\bibinfo{person}{Jon Postel}.} \bibinfo{year}{1981}\natexlab{}.
\newblock \bibinfo{title}{Transmission control protocol}.
\newblock \bibinfo{howpublished}{RFC 793}.
\newblock


\bibitem[Raffelt et~al\mbox{.}(2009)]%
        {dynamicAutomataLearning}
\bibfield{author}{\bibinfo{person}{Harald Raffelt}, \bibinfo{person}{Maik Merten}, \bibinfo{person}{Bernhard Steffen}, {and} \bibinfo{person}{Tiziana Margaria}.} \bibinfo{year}{2009}\natexlab{}.
\newblock \showarticletitle{Dynamic testing via automata learning}.
\newblock \bibinfo{journal}{\emph{International Journal on Software Tools for Technology Transfer (STTT)}} \bibinfo{volume}{11}, \bibinfo{number}{4} (\bibinfo{year}{2009}), \bibinfo{pages}{307--324}.
\newblock
\urldef\tempurl%
\url{https://doi.org/10.1007/s10009-009-0120-7}
\showDOI{\tempurl}


\bibitem[Rasool et~al\mbox{.}(2019)]%
        {RN15}
\bibfield{author}{\bibinfo{person}{Abdullah Rasool}, \bibinfo{person}{Greg Alpár}, {and} \bibinfo{person}{Joeri de Ruiter}.} \bibinfo{year}{2019}\natexlab{}.
\newblock \showarticletitle{State machine inference of QUIC}.
\newblock \bibinfo{journal}{\emph{ArXiv}} (\bibinfo{year}{2019}).
\newblock


\bibitem[Reen and Rossow(2020)]%
        {reen2020dpifuzz}
\bibfield{author}{\bibinfo{person}{Gaganjeet~Singh Reen} {and} \bibinfo{person}{Christian Rossow}.} \bibinfo{year}{2020}\natexlab{}.
\newblock \showarticletitle{DPIFuzz: a differential fuzzing framework to detect DPI elusion strategies for QUIC}. In \bibinfo{booktitle}{\emph{Annual Computer Security Applications Conference (ACSAC)}}. \bibinfo{pages}{332--344}.
\newblock


\bibitem[Rescorla(2018)]%
        {TLS1.3RFC}
\bibfield{author}{\bibinfo{person}{Eric Rescorla}.} \bibinfo{year}{2018}\natexlab{}.
\newblock \bibinfo{title}{The transport layer security ({TLS}) protocol version 1.3}.
\newblock \bibinfo{howpublished}{RFC 8446}.
\newblock


\bibitem[Rescorla and Korver(2003)]%
        {RFC3552}
\bibfield{author}{\bibinfo{person}{Eric Rescorla} {and} \bibinfo{person}{Brian Korver}.} \bibinfo{year}{2003}\natexlab{}.
\newblock \bibinfo{booktitle}{\emph{Guidelines for writing RFC text on security considerations}}.
\newblock \bibinfo{type}{{T}echnical {R}eport}.
\newblock


\bibitem[Schumilo et~al\mbox{.}(2022)]%
        {nyxnet}
\bibfield{author}{\bibinfo{person}{Sergej Schumilo}, \bibinfo{person}{Cornelius Aschermann}, \bibinfo{person}{Andrea Jemmett}, \bibinfo{person}{Ali Abbasi}, {and} \bibinfo{person}{Thorsten Holz}.} \bibinfo{year}{2022}\natexlab{}.
\newblock \showarticletitle{{Nyx-Net}: Network Fuzzing with Incremental Snapshots}. In \bibinfo{booktitle}{\emph{European Conference on Computer Systems (EuroSys)}}.
\newblock
\showISBNx{9781450391627}
\urldef\tempurl%
\url{https://doi.org/10.1145/3492321.3519591}
\showDOI{\tempurl}


\bibitem[Somorovsky(2016)]%
        {somorovsky2016systematic}
\bibfield{author}{\bibinfo{person}{Juraj Somorovsky}.} \bibinfo{year}{2016}\natexlab{}.
\newblock \showarticletitle{Systematic fuzzing and testing of TLS libraries}. In \bibinfo{booktitle}{\emph{ACM SIGSAC Conference on Computer and Communications Security (CCS)}}. \bibinfo{pages}{1492--1504}.
\newblock


\bibitem[Sun et~al\mbox{.}(2023)]%
        {in-context-few-short-learning-2}
\bibfield{author}{\bibinfo{person}{Simeng Sun}, \bibinfo{person}{Yang Liu}, \bibinfo{person}{Dan Iter}, \bibinfo{person}{Chenguang Zhu}, {and} \bibinfo{person}{Mohit Iyyer}.} \bibinfo{year}{2023}\natexlab{}.
\newblock \showarticletitle{How does in-context learning help prompt tuning?}
\newblock \bibinfo{journal}{\emph{arXiv preprint arXiv:2302.11521}} (\bibinfo{year}{2023}).
\newblock


\bibitem[Tappler et~al\mbox{.}(2017)]%
        {RN51}
\bibfield{author}{\bibinfo{person}{Martin Tappler}, \bibinfo{person}{Bernhard~K Aichernig}, {and} \bibinfo{person}{Roderick Bloem}.} \bibinfo{year}{2017}\natexlab{}.
\newblock \showarticletitle{Model-based testing IoT communication via active automata learning}. In \bibinfo{booktitle}{\emph{IEEE International conference on software testing, verification and validation (ICST)}}. \bibinfo{pages}{276--287}.
\newblock
\showISBNx{1509060316}


\bibitem[Thomson and Turner(2021)]%
        {9001}
\bibfield{author}{\bibinfo{person}{Martin Thomson} {and} \bibinfo{person}{Sean Turner}.} \bibinfo{year}{2021}\natexlab{}.
\newblock \bibinfo{title}{{Using TLS to secure QUIC}}.
\newblock \bibinfo{howpublished}{RFC 9001}.
\newblock


\bibitem[Vaandrager(2017)]%
        {modelLearning}
\bibfield{author}{\bibinfo{person}{Frits Vaandrager}.} \bibinfo{year}{2017}\natexlab{}.
\newblock \showarticletitle{Model Learning}.
\newblock \bibinfo{journal}{\emph{Commun. ACM}} \bibinfo{volume}{60}, \bibinfo{number}{2} (\bibinfo{date}{Jan} \bibinfo{year}{2017}), \bibinfo{pages}{86–95}.
\newblock
\showISSN{0001-0782}
\urldef\tempurl%
\url{https://doi.org/10.1145/2967606}
\showDOI{\tempurl}


\bibitem[Walkinshaw and Bogdanov(2013)]%
        {lts_diff}
\bibfield{author}{\bibinfo{person}{Neil Walkinshaw} {and} \bibinfo{person}{Kirill Bogdanov}.} \bibinfo{year}{2013}\natexlab{}.
\newblock \showarticletitle{Automated comparison of state-based software models in terms of their language and structure}.
\newblock \bibinfo{journal}{\emph{ACM Transactions on Software Engineering and Methodology (TOSEM)}} \bibinfo{volume}{22}, \bibinfo{number}{2} (\bibinfo{year}{2013}), \bibinfo{pages}{1--37}.
\newblock


\bibitem[Xia et~al\mbox{.}(2024)]%
        {xia2024fuzz4all}
\bibfield{author}{\bibinfo{person}{Chunqiu~Steven Xia}, \bibinfo{person}{Matteo Paltenghi}, \bibinfo{person}{Jia Le~Tian}, \bibinfo{person}{Michael Pradel}, {and} \bibinfo{person}{Lingming Zhang}.} \bibinfo{year}{2024}\natexlab{}.
\newblock \showarticletitle{Fuzz4all: Universal fuzzing with large language models}. In \bibinfo{booktitle}{\emph{IEEE/ACM International Conference on Software Engineering (ICSE)}}. \bibinfo{pages}{1--13}.
\newblock


\end{thebibliography}
\appendix \label{sec:appendix}

%-------------------------------------------------------------------------------

\section*{Appendix}
\section{Symbolization with LLMs}\label{sec:llm_symbol_study_appendix}
%-------------------------------------------------------------------------------

Recently, language models (LLMs) have been investigated as potential tools for supporting fuzzing~\cite{chatafl, xia2024fuzz4all, llm-zero-shot-deep-lib, llvm-few-short-bug-repro}. In compliance testing with active learning, we explore the utility of LLMs to automate the manual task of examining a protocol specification to generate the input and output symbols for the QUIC \learner{}. We perform an evaluation on OpenAI's GPT-4o (the most advanced model at the time of writing) using the prompt engineered in Figure~\ref{fig:llm-prompt}. Inspired by~\cite{chatafl}, to ensure the LLM model always returns the symbols in a consistent format, we employ \textit{in-context few-shot learning}~\cite{in-context-few-short-learning-1, in-context-few-short-learning-2}, a prompt engineering technique that helps the LLM model to understand the desired pattern of the output based on given examples. In this evaluation, we ask the LLM to return \textit{only} valid messages that are exchanged during a QUIC handshake. The symbols from the responses can be grouped into 2 categories: i)~valid symbols---valid messages in the desired format; and ii)~invalid symbols---either invalid messages (incorrect packet and frame combination) or symbols that do not align with the format we requested. We sampled 50 responses from the LLM and combined them into one symbol set as shown in Figure~\ref{fig:llm_symbol_eval}. We measure the number of misses and hallucination rates. Number of misses refers to the generation of missing information based on the ground truth, and Hallucination refers to the generation of untruthful information~\cite{chen2023can}.

From our evaluation, we see that the answers given by the LLM is not guaranteed and consistent. The ground truth, extracted from manually inspecting the specification, consists of 25 symbols; however, the LLM only generates 8.6 symbols on average (3 symbols are given as part of the prompt). This results in a 69\% missing rate on average as shown in Table~\ref{tab:symbolsMissingRate}. In addition, as shown in Table~\ref{tab:hullucination-rate} the symbols sampled across 50 queries have an average of 18.6\% hallucination rate. Therefore, we do not rely on symbol generation using an LLM model because it requires multiple queries to get all the correct symbols and, subsequently, an expert with protocol-specific knowledge to examine the symbols to eliminate undesirable invalid symbols.

\begin{figure}[ht]
  \centering
  \begin{roundedtextbox}[title=Prompt]
    \textsf{Instruction}: According to RFC 9000 (QUIC specification), list all the packet and frame type combinations that can exist during a QUIC handshake. \\
    
    \textsf{Desired Format}: \\
    
    \textbf{Shot-1}: \\
    For the QUIC protocol, the Initial packet with CRYPTO frame carrying a Client Hello TLS message will be: \\
    Initial\_Client\_Hello \\
    
    \textbf{Shot-2}: \\
    For the QUIC protocol, the Handshake packet with a PING frame will be: \\
    Handshake\_Ping \\
    
    \textbf{Shot-3}: \\
    For the QUIC protocol, the 1-RTT packet with CRYPTO frame carrying a Finished TLS message frame will be: \\
    Handshake\_Finished
  \end{roundedtextbox}
  \vspace{-3mm}
  \caption{An example prompt to automatically symbolize all valid messages during a QUIC handshake.    }
  \label{fig:llm-prompt}
\end{figure}

\begin{figure}[t!]
    \centering
    \includegraphics[width=0.95\columnwidth]{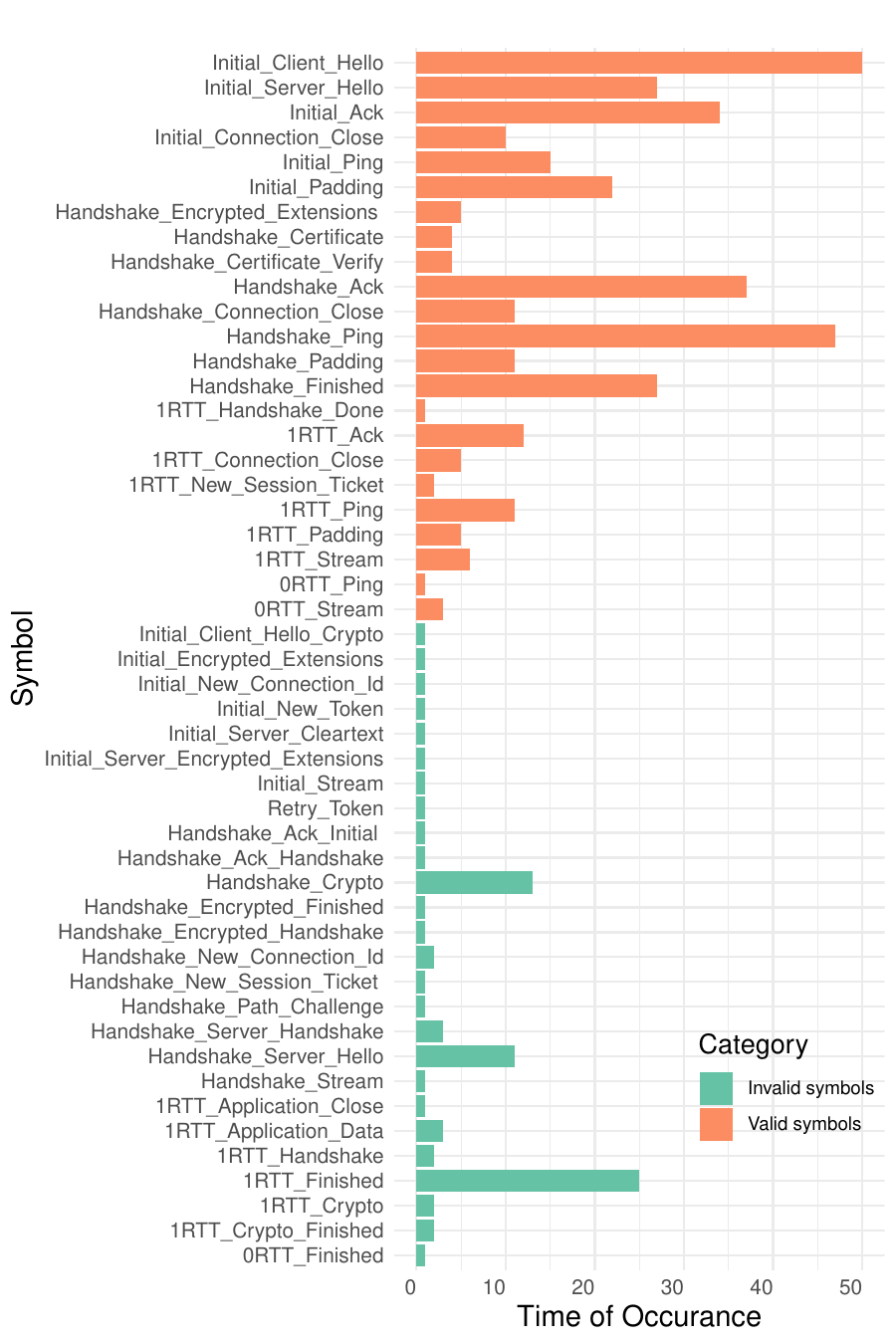}
    \caption{Symbols in 50 answers sampled from OpenAI GPT-4o model using the prompt shown in Figure~\ref{fig:llm-prompt}.}
    \label{fig:llm_symbol_eval}
    \vspace{-2mm}
\end{figure}

\begin{table}[h!]
\centering
\caption{Symbol missing rate in 50 LLM queries compared to the ground truth extracted from RFC9000 by a domain expert.}
\vspace{-2mm}
\label{tab:symbolsMissingRate}
\resizebox{1.0\columnwidth}{!}{%
\begin{tabular}{lccr}
\toprule
\textbf{Symbol (Ground Truth)}                           & \textbf{Number of Occurances}    & \textbf{Number Missed}                 & \textbf{Missing Rate(\%)}       \\ \midrule
Initial\_Client\_Hello           & 50                   & 0                           & 0\%                       \\
Initial\_Server\_Hello           & 27                   & 23                          & 46\%                      \\
Initial\_Ack                     & 34                   & 16                          & 32\%                      \\
Initial\_Connection\_Close       & 10                   & 40                          & 80\%                      \\
Initial\_Ping                    & 15                   & 35                          & 70\%                      \\
Initial\_Padding                 & 22                   & 28                          & 56\%                      \\
Handshake\_Encrypted\_Extensions & 5                    & 45                          & 90\%                      \\
Handshake\_Certificate           & 4                    & 46                          & 92\%                      \\
Handshake\_Certificate\_Verify   & 4                    & 46                          & 92\%                      \\
Handshake\_Ack                   & 37                   & 13                          & 26\%                      \\
Handshake\_Connection\_Close     & 11                   & 39                          & 78\%                      \\
Handshake\_Ping                  & 47                   & 3                           & 6\%                       \\
Handshake\_Padding               & 11                   & 39                          & 78\%                      \\
Handshake\_Finished              & 27                   & 23                          & 46\%                      \\
1RTT\_Handshake\_Done            & 1                    & 49                          & 98\%                      \\
1RTT\_Ack                        & 12                   & 38                          & 76\%                      \\
1RTT\_Connection\_Close          & 5                    & 45                          & 90\%                      \\
1RTT\_New\_Session\_Ticket       & 2                    & 48                          & 96\%                      \\
1RTT\_Ping                       & 11                   & 39                          & 78\%                      \\
1RTT\_Padding                    & 5                    & 45                          & 90\%                      \\
1RTT\_Stream                     & 6                    & 44                          & 88\%                      \\
0RTT\_Ping                       & 1                    & 49                          & 98\%                      \\
0RTT\_Stream                     & 3                    & 47                          & 94\%                      \\
Retry                            & \textbf{0}                   & 50                          & \multicolumn{1}{r}{100\%} \\
1RTT\_New\_Connection\_Id        & \textbf{0}                    & 50                          & 100\% \\ \midrule
                                 &                       & \textbf{Average}                   & \textbf{69\%} \\ \bottomrule
\end{tabular}%
}
\end{table}

\begin{table}[]
\centering
\caption{The hallucination rate of the symbols sampled across 50 queries.}
\vspace{-2mm}
\label{tab:hullucination-rate}
\resizebox{1.0\columnwidth}{!}{%
\begin{tabular}{lcccr}
\toprule 

\begin{tabular}{@{}l@{}}\textbf{Number of} \\ \textbf{Queries} \end{tabular}& \begin{tabular}{@{}c@{}} \textbf{Number of} \\ \textbf{Symbols Sampled} \end{tabular} & \begin{tabular}{@{}c@{}} \textbf{Number of} \\ \textbf{Correct Symbols} \end{tabular} & \begin{tabular}{@{}c@{}} \textbf{Number of} \\ \textbf{Invalid Symbols} \end{tabular} & \begin{tabular}{@{}r@{}} \textbf{Hallucination} \\ \textbf{Rate(\%)} \end{tabular} \\ \midrule
50                & 430                       & 350                       & 80                        & 18.6     
\\ \bottomrule
\end{tabular}%
}
\end{table}

\subsection{Extended Material on Automatically Learning a Behavior Model  and Symbolization (\learner{})}\label{sec:symb_appendix}
We described the general role of a learner in Section~\ref{sec:protocolStateFuzzing}---recall, the \learner{} generates input sequences from a set of input symbols. We curated and defined symbols for protocol parameters and messages, parameterized by time, for QUIC. These are summarized in Table~\ref{tab:symbolDictionary} in the Appendix. Notably, to increase the readability of the models, these symbols are represented using their acronyms. Specifically, for input test cases, we included symbols:

\begin{itemize}[itemsep=1pt,parsep=1pt,topsep=1pt,labelindent=0pt,leftmargin=5mm]
    \item To construct \textit{all} valid QUIC messages in a handshake, including \pmsg{Ping} and \pmsg{ConnectionClose} messages, to increase the learning coverage as these messages can exist in both \pmsg{Initial} and \pmsg{Handshake} packets~\cite[Section 17]{9000}. 
    \item To allow constructing \textit{invalid} messages violating the specification. These messages carry one of the following: i)~incorrect header fields (e.g. Connection ID); ii)~not permitted frame structures (e.g. frame without any content); and iii)~frame with incorrect content (e.g. Invalid Certificate). 
\end{itemize}

\noindent While, for QUIC protocol responses or output symbols, we include symbols:

\begin{itemize}[itemsep=1pt,parsep=1pt,topsep=1pt,labelindent=0pt,leftmargin=5mm]
    \item To construct \textit{all} valid QUIC client response messages. 
    \item To elicit the status (connection availability) of the \sut{} which would otherwise be hidden from the \learner{}. This addition is important to detect DoS attacks (such as \textbf{M}-18 in Table~\ref{tab:bug-table-summary}) and uncover non-compliant behaviors of the \sut{}; process a message where it should not, for example, \pmsg{initialConnectionClose} after employing the Handshake encryption key~\cite[Section 4.9]{9001}.
\end{itemize}

So, the crafted input symbols include \textit{all} valid QUIC client messages described in Section~\ref{sec:quicHandshake}. We also added \pmsg{Ping} and \pmsg{ConnectionClose} messages to increase the learning coverage as these messages can exist in both \pmsg{Initial} and \pmsg{Handshake} packets~\cite[Section 17]{9000}. Additionally, we include \textit{12} invalid QUIC messages that violate the specifications~\cite{9000,9001,TLS1.3RFC}: i)~\pmsg{initialClientHello-invldACK}; ii)~\pmsg{{handshakeEmptyCertificate}}; iii)~\pmsg{{handshakeInvalidCertificate}}; iv)~\pmsg{{InvalidNewConnectionID}}; v)~\pmsg{initialNoFrame}; vi)~\pmsg{initialUnexpectedFrameType}; vii)~\pmsg{handshakeNoFrame}; viii)~\pmsg{handshakeUnexpectedFrameType}; ix)~\pmsg{0rttNoFrame}; x)~\pmsg{0rttUnexpectedFrameType}; xi)~\pmsg{0rttFinished}; and xii)~\pmsg{0rttACK}. We also crafted two input symbols for \mapper{} configuration settings for the \learner{} to select during learning: i)~\pmsg{[RemovePaddingFromInitialPackets]}; and ii)~\pmsg{[ChangeDestination-ConnectionID-Original]}. A detailed description of each input is included in Table~\ref{tab:symbolDictionary}.

The output symbols are message types a QUIC protocol will respond with, these include the messages described in Section~\ref{sec:quicHandshake}, \pmsg{PingACK} and \pmsg{ConnectionClose}. Importantly, we include two output symbols to elicit the status of the \sut{} which would otherwise be hidden from the \learner{}. We check the \sut{} status by sending a ping message after sending a \pmsg{ConnectionClose} message carried by either an \pmsg{Initial} or a \pmsg{Handshake} packet. The \sut{} is considered alive if it acknowledges the ping. The symbols are enclosed in angle brackets. \pmsg{{$<$ConnectionClose$>$}} indicates the \sut{} has closed the connection, and \pmsg{{$<$ConnectionActive$>$}} denotes that the connection with the \sut{} is still active. This addition is important for testing the \textsf{SUT} state for conditions where it should not process a message, for example, \pmsg{initialConnectionClose} or \pmsg{handshakeConnectionClose} messages as stated in \cite[Section 4.9]{9001}. A detailed description of each output is included in Table~\ref{tab:symbolDictionary}.

\section{Demonstrating a Valid Behavior Analysis}
\label{appendix:valid_demo}
\begin{figure*}[t!]
    \centering
    \includegraphics[width=1.0\linewidth]{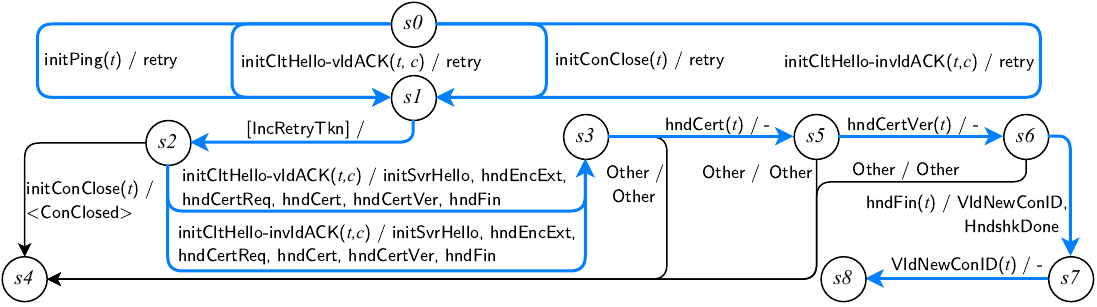}
    \vspace{-4mm}
    \caption{Optimized learned model of a Ngtcp2 server with the \retryClientAuthentication{} configuration, generated from a \learner{} with the symbols required for \retryClientAuthentication{} to generate a simplified model for illustration. Here, blue edges are the valid path to complete a \retryClientAuthentication{} QUIC handshake.}
    \vspace{-2mm}
    \label{fig:Ngtcp2Optimised}
\end{figure*}

In Figure~\ref{fig:Ngtcp2Optimised}, \emph{s0} denotes the state the server is fully initialized and waiting for incoming connections. The client first sends an \pmsg{{initial-ClientHello-validACK}} and receives a \pmsg{{retry}} from the server for client address validation, then transitions to \emph{s1}. With the \pmsg{{[IncludeRetryToken]}} input at \emph{s1}, the client is configured to start including the received \pmsg{RetryToken} in all its following \pmsg{Initial} packets. From \emph{s2} to \emph{s3}, the client sends an \pmsg{{initialClientHello-validACK}} that includes the received \pmsg{RetryToken}. The Ngtcp2 server verifies the \pmsg{RetryToken} and responds with \pmsg{{initialServerHello}}, \pmsg{{handshakeEncryptedExtension}}, \pmsg{{handshakeCertificateRequest}}, \pmsg{{handshakeCertificate}}, \pmsg{{handshakeCertificateVerify}}, \pmsg{{handshakeFinished}} and \pmsg{{ValidNewConnectionID}} messages. Importantly, the \pmsg{{handshakeCertificateRequest}} message indicates to the client it must present a certificate for authentication to proceed.

Subsequently, the client sends \pmsg{{handshakeCertificate}} to the server at \emph{s3}, followed by \pmsg{{handshakeCertificateVerify}}. At this point, the server verifies the client certificate, and the handshake transitions to \emph{s6}. The client sends  \pmsg{{handshakeFinished}} and the server validates the message to ensure that the previous handshake messages have not been modified. The handshake transitions to \emph{s7} after the server responds with \pmsg{{ValidNewConnectionID}} and \pmsg{{HandshakeDone}} messages. Now the QUIC handshake is considered confirmed. The handshake ends at \emph{s8} after the client completes the exchange of the \pmsg{{ValidNewConnectionID}} message that carries several \pval{new connection IDs} that can be used for the established connection. Notably, the handshake can still proceed despite the client sending \pmsg{{initialClientHello-invalidACK}} at \emph{s2}, which contains an invalid \pmsg{Initial} ACK to acknowledge the server's \pmsg{{initialServerHello}} message. This is because the server recognizes and drops the invalid \pmsg{Initial} ACK. But, it continues with the handshake process when it receives a valid \pmsg{Handshake} ACK from the client, acknowledging the \pmsg{Handshake} packets correctly. \textit{This is an example of the expected handshake flow according to the specification.}

%%%%%%%%%%%%%%%%%%%%%%%%% Quiche Authentication Bypass start &&&&&&&&&&&&&&&&&&&&&&&&&&&&&&&&&&&&

\section{Additional Case Studies}
\label{appendix:bug_case_studies}
We provide additional case studies in this section, covering more memory corruption bugs and logical flaws discovered by \name{}.

%%%%%%%%%%%%%%%%%%%%%%%%% Memory-corruption bug  start &&&&&&&&&&&&&&&&&&&&&&&&&&&&&&&&&&&&
\vspace{1mm}
\subsection{Memory-corruption bugs: Server crashes} \label{sec:PicoquicAppendix}
\label{appendix:memory_case_studies}

\begin{mdframed}[backgroundcolor=black!5,rightline=false,leftline=false,topline=false,bottomline=false,roundcorner=1mm,everyline=false]
We discovered \softwareBugs{} memory-corruption bugs. The inputs for reproducing each bug as well as a detailed description of each bug are available at our open-source \name{} code repository on GitHub~\textsf{\href{https://github.com/QUICTester}}~\cite{repo-inputs}. Here, in addition to Section~\ref{sec:Picoquic}, we detail three additional memory-corruption bugs that can result in DoS attacks.
\end{mdframed}
\vspace{3mm}

%%%%%%%%%%%%%% QUINN %%%%%%%%%%%%%%
\label{sec:nullPointerDeferenceInQuinnAppendix}
\noindent\Circled{M-8} \textbf{Null Pointer Dereference in Quinn.}~When testing Quinn, our crash logger detected crashes when handling \pmsg{hndUnxpFrType}. These crashes arise from Quinn \textit{panicking} when it attempts to unwrap a  {\ttfamily\hyphenchar\font=`\-None} value after matching an unexpected frame to the {\ttfamily\hyphenchar\font=`\-Type} enum. 

\vspace{1mm}
\noindent\textit{\Circled{Impact}~}This allows an attacker to perform a DoS attack using a malformed packet (\pmsg{hndUnxpFrType}). Notably, the bug exists in both the server and client implementations since they share the same library. We responsibly reported this vulnerability to the Quinn developers. This vulnerability was assigned \textsf{CVE-2023-42805} with high severity and patched.

%%%%%%%%%%%%%%%%%%%%%%%%%%% Neqo  %%%%%%%%%%%%%%%%%%%%%%%%%%%%%%%%%%%%
\vspace{2mm}
\noindent\Circled{M-2} \textbf{Null Pointer Dereference in Neqo.~}Upon looking at the \learner{} logs, it was found that the Neqo server crashed with an assertion error in cases where the selected input sequence contains an \pmsg{{initialConnectionClose}} message that precedes an \pmsg{{initialClientHello}} message. This assertion is also guaranteed to occur when the input sequence includes the \pmsg{{initialConnectionClose}} message but lacks the \pmsg{{initialClientHello}} message. Based on our findings, it appears that the server attempts to respond with a \pmsg{{ConnectionClose}} message. However, it cannot obtain the connection's primary path when creating the message. This happens because the server will only set a primary path for that connection when it receives and processes an \pmsg{{initialClientHello}} message. This finding demonstrated that the \learner{} logs are helpful in detecting memory-corruption bugs that are not directly shown in the learned models. 

\vspace{2mm}
\noindent\textit{\Circled{Impact}~}This vulnerability allows an attacker to launch a denial of service (DoS) attack on Neqo servers by sending a single \pmsg{{initialConnectionClose}} input at the start of a connection establishment.

\vspace{2mm}
\noindent\Circled{M-3} \textbf{Limited connections due to a hardcoded value.~}The Neqo server in all configurations can only accept at most 32,767 connections, including closed connections. After the 32,767\textsuperscript{th} connection, the server crashes with an assertion error when the variable of \texttt{PRDescIdentity} data type is storing a value that is equal to \texttt{int\_16\_max()} at the beginning of the \texttt{PD\_GetUniqueIdentity()} function. This function creates a unique identity for each connection, and each identity is assigned a unique identity number. The unique identity number starts from 0 and then increases by 1 for every new unique identity created. To ensure the identity is unique, the server uses a variable to track the most recent unique identity number. In the \texttt{PD\_GetUniqueIdentity()} function, if the current unique identity number is equal to int\_16\_max(), it will stop the unique identity creation and raise an assertion error. As described in Mozilla's documentation, the data type, \texttt{PRIntn}/\texttt{PRDescIdentity}, that is used to store the unique identity number is guaranteed to be at least 16 bits, but the architecture that runs the server can define it to be wider, such as 32 bits or 64 bits. However, due to the hard-coded comparison value, int\_16\_max() in the assertion statement, our architecture that defined \texttt{PRDescIdentity} to 32 bits which can actually create up to 2,147,483,647 unique identities, will still crash after the 32767\textsuperscript{th} connection. This may affect the performance of the Neqo server because, with the hard-coded value in the assertion, it is guaranteed to crash on the 32,768\textsuperscript{th} connection, no matter the data type used by the architecture. 

\vspace{1mm}
\noindent\textit{\Circled{Impact}~}This allows an attacker to perform a DoS attack by establishing more than 32,767 QUIC connections with the server. Both \textbf{M-2} and \textbf{M-3} were fixed by Neqo developers. Notably, the developers stated these vulnerabilities only affect the server-side implementation as they currently focus on client implementation.

%%%%%%%%%%%%%%%%%%%%%%%%% Logic bug  start &&&&&&&&&&&&&&&&&&&&&&&&&&&&&&&&&&&&
\subsection{Logical Flaws: Unexpected Behaviors}
\label{appendix:bug_case_studies}

\begin{mdframed}[backgroundcolor=black!5,rightline=false,leftline=false,topline=false,bottomline=false,roundcorner=1mm,everyline=false]
We discovered \logicBugs{} logical flaws. The inputs for reproducing each bug as well as a detailed description of each bug are available at our open-source \name{} code repository on GitHub~\textsf{\href{https://github.com/QUICTester}}~\cite{repo-inputs}.  Here, we detail two more logical flaws.
\end{mdframed}

\vspace{2mm}
\noindent\Circled{L-2} \textbf{Incorrect method of emptying the re-transmission queue.} Because Pquic is built on top of the older Picoquic library, it shared the issue \textbf{M-4} discussed in Section~\ref{sec:picoquic_memory} with Picoquic. However, unlike Picoquic, which attempts to access a \texttt{NULL} pointer to obtain the encryption key for re-transmission, Pquic will always have access to the encryption keys because it never discards them---see \textbf{S-10}.

\vspace{1mm}
\noindent\textit{\Circled{Impact}~}This behaviour causes the server to re-transmit acknowledged messages to the peer, unnecessarily increasing network traffic and reducing the utility of network bandwidth.

\vspace{2mm}
\noindent\Circled{L-3} \textbf{Infinite loop when processing frame type} \texttt{0xFF}.~When the PQUIC server processes a packet carrying a \texttt{0xFF} frame type, the server always gets stuck in a loop that attempts to match \texttt{0xFF}, an invalid frame type, with the expected frame type. This issue is specific to the \texttt{0xFF} frame type and does not happen with other invalid frame types.

\vspace{1mm}
\noindent\textit{\Circled{Impact}~}Because PQUIC is running on a single thread, getting stuck in an infinite loop causes the PQUIC server to become unavailable to serve any client until the server administrator manually restarts it. This allows an attacker to perform a DoS attack on the server using the message described above.

\begin{figure}[t!]
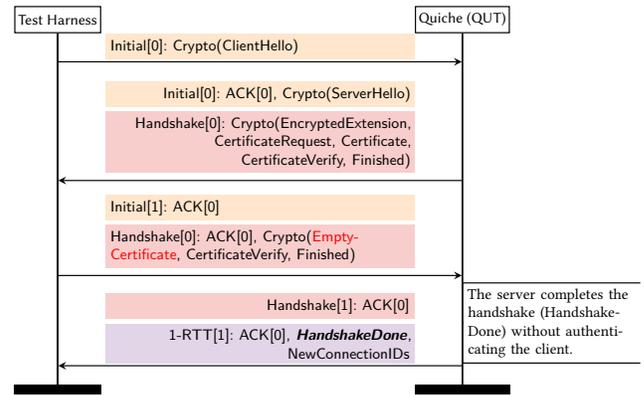

%\resizebox{1\columnwidth}{!}{
\centering
 \resizebox{1.0\columnwidth}{!}{
    \begin{msc}[instance distance = 6.6cm, draw frame = none, environment distance = 0, instance width = 1.1cm, msc keyword=, head top distance=0cm]{} %
    \centering
    \declinst{client}{}{\textsf{\mapper{}}}
    \declinst{server}{}{\textsf{Quiche (\sut{})}}
    \mess[label distance = 0.3ex, align=left]{\colorbox{BackgroundOrange}{\parbox{6.3cm}{\raggedright \pmsg{{Initial[0]: Crypto(ClientHello)}}}}}{client}{server}
    \nextlevel[5]
    \mess[label distance = 0.3ex, align=right, draw]{
    \colorbox{BackgroundOrange}{\parbox{6.3cm}{\raggedleft \pmsg{{Initial[0]: ACK[0], Crypto(ServerHello)}}}}\\[0.5ex]
    \colorbox{BackgroundRed}{\parbox{6.3cm}{\raggedleft \pmsg{{Handshake[0]: Crypto(EncryptedExtension,\\
    CertificateRequest, Certificate, CertificateVerify, Finished)}}}}\\[0.5ex]
    }{server}{client}
    \nextlevel[4]
    \mess[label distance = 0.3ex, align=left]{
    \colorbox{BackgroundOrange}{\parbox{6.3cm}{\raggedright \pmsg{{Initial[1]: ACK[0]}}}}\\[0.5ex]
    \colorbox{BackgroundRed}{\parbox{6.3cm}{\raggedright \pmsg{{Handshake[0]: ACK[0], Crypto(\textcolor{red}{Empty-\\Certificate}, CertificateVerify, Finished)}}}}\\[0.5ex]
    }{client}{server}
    \nextlevel[2]
    \msccomment[%
  msccomment distance=0cm,
  side=right,
  /msc/every msccomment/.append style={
   text width=3.5cm
  }
  ]{The server completes the handshake (HandshakeDone) without authenticating the client.}{server}.
    \nextlevel[1.8]
    \mess[label distance = 0.3ex, align=right]{
    \colorbox{BackgroundRed}{\parbox{6.3cm}{\raggedleft \pmsg{{Handshake[1]: ACK[0]}}}}\\[0.5ex]
    \colorbox{BackgroundPurple}{\parbox{6.3cm}{\raggedleft \pmsg{{1-RTT[1]: ACK[0], \textbf{\textit{HandshakeDone}},\\
    NewConnectionIDs}}}}
    }{server}{client}
    \end{msc}
    }
    \caption{Client authentication bypass in Quiche. The invalid (EmptyCertificate) message is shown in \textcolor{red}{red text}.}
    \label{fig:quicheClientAuthBypass}
%}
\end{figure}

%%%%%%%%%%%%%%%%%%%%%%%%%%%%%%%%%%%%%%%%%

\section{Reference Models}\label{sec:refmode-appendix}
As explained in Section~\ref{sec:discussion}, we have curated 11 reference models from our experiment. These reference models are FSMs with no specification violations. Our library has at least one reference model for each security configuration we tested that developers can use with \name{}. We have curated 11 reference models (3 for Basic, 2 for Retry, 2 for ClientAuth, 1 for RetryClientAuth and 3 for PSK). The main differences between the reference models in the same security configuration but from different vendors are due to the variations in the interpretation and implementation of the response to the \pmsg{initialPing} message sent by a client as the first message. For example, Ngtcp2 server drops the first \pmsg{initialPing} message from clients, while the Quicly server responds to the \pmsg{initialPing}. As we discussed in Section~\ref{sec:ambiguity}, the current specification does not explicitly state the expected behavior of a server to a first packet received without a CRYPTO frame. In our reference models, we consider both implementations as adhering to the specification. All the reference models can be found on our public GitHub repository \href{https://github.com/QUICtester}{https://github.com/QUICtester}.

\section{\name{} Implementation Effort}
\label{sec:impefffort_appendix}
\noindent We summarize our implementation effort in Table~\ref{tab:code-lines}.

\begin{table}[h!]
\centering
\caption{Extensions made to implement QUICTester.}
\vspace{-2mm}
\label{tab:code-lines}
\centering
\begin{tabular}{lcc}
\toprule
\textbf{Component}   & \textbf{Library} & \textbf{Lines of Code} \\ \toprule
Learner               & LearnLib   & 627           \\ \midrule
Mapper                & Aioquic & 3560          \\ \midrule
Optimizer             & -       & 416           \\ \midrule
Differential Analyzer & LTSDiff & 719          \\ \bottomrule
\end{tabular}
\end{table}

\begin{figure*}[t]
 \resizebox{1\textwidth}{!}{
    \centering
    \includegraphics[trim=1cm 1.2cm 1cm 1.2cm, clip, height=0.5\textwidth, width=1.0\linewidth]{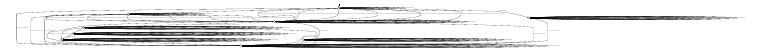}}
    \caption{The learned model of Quinn---notably the QUIC implementation with the valid FSM---in the most simple, \basic{} configuration before optimization. \textcolor{blue}{This illustration, whilst not fully legible, is provided to show an example of the complexity created in even the most basic security configuration for a learned model before optimization.} The model after using our \opt{} is shown in Figure~\ref{fig:optimizedLearnedModel}.
}
    \label{fig:learnedModel}
\end{figure*}

\begin{figure*}[h]
 \resizebox{1.0\textwidth}{!}{
    \includegraphics[trim=0.5cm 1cm 0.5cm 1cm, clip, height=0.7\textwidth, width=1.5\linewidth]{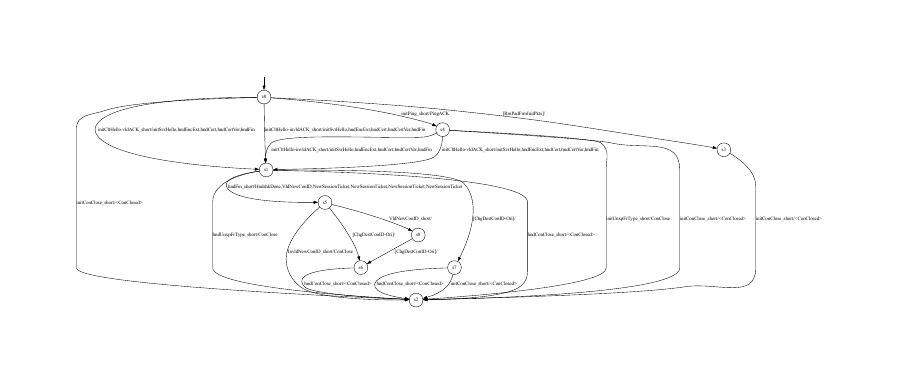}}
    \caption{The learned model of Quinn \basic{} after employing our \opt{} (compared with Figure~\ref{fig:learnedModel} generated prior to simplification, there are far fewer edges, greatly improving \textit{interpretability} and the task of model analysis).}
    \label{fig:optimizedLearnedModel}
\end{figure*}

\begin{figure*}[t!]
    \centering
    \includegraphics[width=1.0\linewidth]{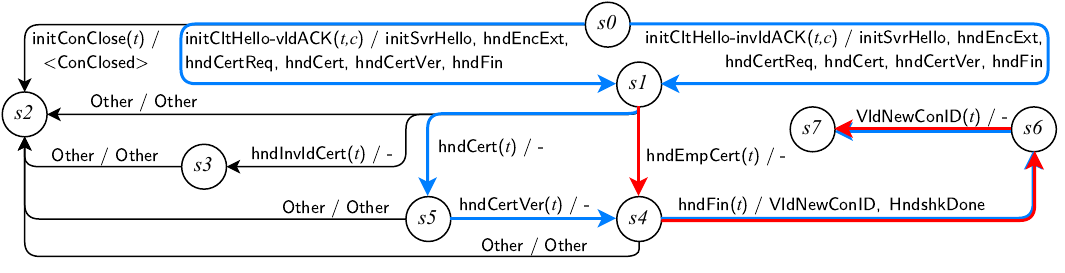}
    \caption{Optimized learned model of a Quiche server with the \clientAuthentication{} configuration. Blue edges show a valid path to complete a QUIC handshake. Red edges demonstrate an invalid path that bypasses the client authentication, but still completes the handshake without errors.}
    \label{fig:quicheClienttAuth}
\end{figure*}

\begin{figure*}[t!]
    \centering
    \includegraphics[width=1.0\linewidth]{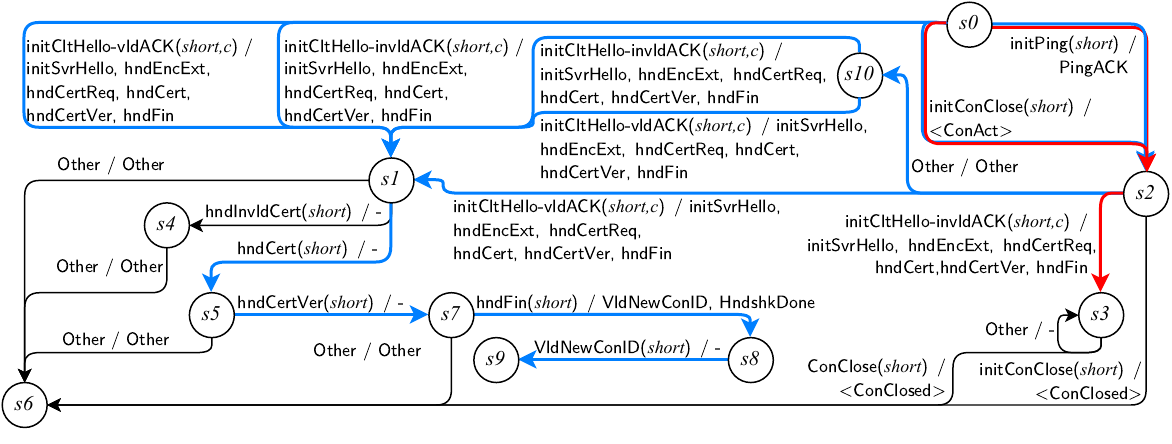}
    \caption{An optimized model from Picoquic with \clientAuthentication{} configuration learned with $t=short$ parameter setting for inputs. Differental analysis with Figure~\ref{fig:picoOptimisedLong} using $t=long$ reveals a software bug exploit.}
    \label{fig:picoOptimisedShort}
\end{figure*}

\begin{figure*}[t!]
    \centering
    \includegraphics[width=1.0\linewidth]{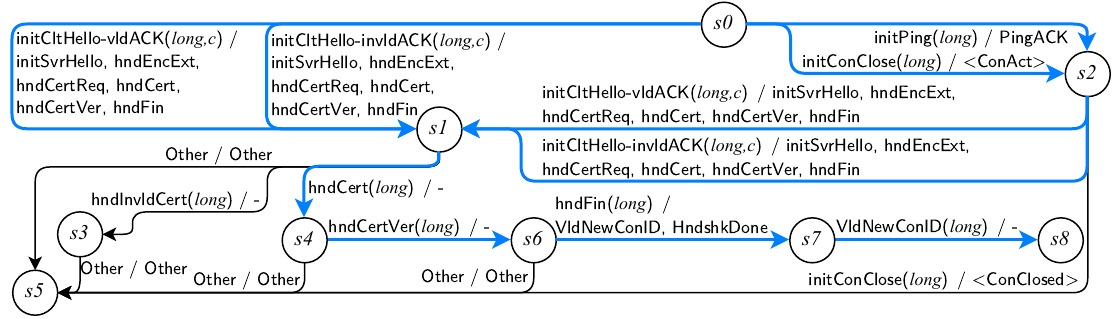}
    \caption{An optimized model from Picoquic with \clientAuthentication{} configuration learned with $t=long$ parameter setting for inputs.}
    \label{fig:picoOptimisedLong}
\end{figure*}
\clearpage

%% Bug-table %%
\begin{table*}[t]
\centering
\caption{Overview of identified faults. In total, \protocolViolation{} specification violations, \softwareBugs{} memory-corruption bugs and \logicBugs{} logical flaws were identified in \implementationcount{} QUIC implementations. ~\textit{\textbf{Specification bugs}} \textbf{(S):} An implemented behavior violates the QUIC specification. \textit{\textbf{Memory-corruption bugs}} \textbf{(M):} An input causing a memory corruption and a server crash. \textit{\textbf{Logical flaws}} \textbf{(L):} Incorrect logic implemented in code produces unexpected behavior. 
}
\label{tab:bug-table}
\resizebox{\textwidth}{!}{%
\begin{tabular}{p{3cm}p{11.5cm}ccp{5.5cm}}
\toprule
\textbf{Server} & \textbf{Fault Description} & \textbf{Type}-ID & \textbf{Disclosed} & \textbf{Most Recent Response to Disclosure}  \\ \toprule
Aioquic & Incorrect handling of packets with unexpected frame type. & \textbf{S}-1 & \checkmark  & Fixed. \\ \midrule

Kwik & CVE-2024-22588: Retention of the unused encryption keys. & \textbf{S}-2 & \checkmark & Fixed.   \\ %\cline{2-5} 
& CVE-2024-22590: Implementation without a TLS state machine. & \textbf{S}-3 & \checkmark & Fixed.  \\ 
& Process CRYPTO frame in a 0-RTT packet. & \textbf{S}-4 & \checkmark & Acknowledged findings.  \\ 
& Exceeds the operating system’s maximum number
of memory mappings for a single process (100,000) when receiving PING frame from 50,000 clients. & \textbf{M}-1 & \checkmark & Acknowledged findings.  \\ \midrule

Lsquic~(Lite Speed) & CVE-2024-25678: Retention of the unused encryption keys (PSK configuration in 1b113d19). & \textbf{S}-5 & \checkmark & Fixed.   \\
 & Incorrect handling of re-transmission, leaving a half-opening connection on the client side (PSK configuration in v4.0.2). & \textbf{L}-1 & \checkmark & Fixed.   \\ \midrule

MsQuic~(Microsoft) & Does not issue its initial\_source\_connection\_id at the correct connection state. This finding is part of the connection management ambiguity discussed in Section~\ref{sec:ambiguity}. & \textbf{S}-6 & \checkmark & Developers acknowledge the findings and plan to propose an amendment to the \textit{QUIC specification} to address the ambiguity. \\ \midrule
% Reported on 18 July 2023. 

Neqo~(Mozilla) & NULL pointer dereference when getting the primary path. & \textbf{M}-2 & \checkmark & 
Fixed. \\
 %\cline{2-5} 
 & Limited connections due to a hardcoded value. & \textbf{M}-3 & \checkmark &  Fixed.\\ \midrule

Picoquic & NULL pointer dereference when getting the encryption keys. & \textbf{M}-4 & \checkmark & Fixed. \\ %\cline{2-5} 
 & Retry token tied to retry\_source\_connection\_id.  & \textbf{S}-7 & \checkmark & Developer stated  the code was written specifically to add an additional constraint to ensure the client follows the specification. Fix: Propose an \textit{amendment to the QUIC specification} to address the ambiguity.\\ \midrule

PQUIC & Invalid original\_destination\_connection\_id. & \textbf{S}-8 & \checkmark & Fixed. \\  
 & Limitless active\_connection\_id\_limit. & \textbf{S}-9 & \checkmark & Fixed. \\ 
& CVE-2024-25679: Retention of the unused encryption keys. & \textbf{S}-10 & \checkmark & Fixed. \\
& Incorrect way of emptying the re-transmission queue. & \textbf{L}-2 & \checkmark & Fixed. \\ 
& NULL pointer dereference when handling removed connection context. & \textbf{M}-5 & \checkmark & Acknowledged findings. \\ 
& Buffer overflow when processing frame type 0x30. & \textbf{M}-6 & \checkmark & Acknowledged findings. \\ 
& Infinite loop when processing frame type 0xFF. & \textbf{L}-3 & \checkmark & Acknowledged findings. \\
& Does not send HANDSHAKE\_DONE after the handshake is confirmed (PSK configuration). & \textbf{S}-11 & \checkmark & Acknowledged findings. \\\midrule

Quiche~(Cloudflare) & Client authentication bypass due to incorrect flag set in Quiche library. & \textbf{S}-12 & \checkmark & Fixed with bug bounty awarded.  \\
 & Incorrect handling of \textbf{all} Initial packets carried in a UDP datagram with a payload size smaller than 1200 bytes. & \textbf{S}-13 & \checkmark & Acknowledged findings.  \\ \midrule

Quiche4j & Concurrent modification exception when discarding closed connections. & \textbf{M}-7 & \checkmark & Acknowledged findings. \\ 
& Limitless active\_connection\_id\_limit. & \textbf{S}-14 & \checkmark & Acknowledged findings. \\
\midrule

Quant & Incorrect handling of an \pmsg{initialPing} message. & \textbf{S}-15 & \checkmark & \multirow{2}{*}{\begin{tabular}{@{}l@{}}
Acknowledged findings. Not fixed \\because the GitHub Repository\\ is no longer under active maintenance. 
\end{tabular}} \\
 & Incorrect handling of \textbf{all} Initial packets carried in a UDP datagram with a payload size smaller than 1200 bytes. & \textbf{S}-16 & \checkmark &  \\  \midrule

Quiwi & Does not close the connection when the number of received NEW\_CONNECTION\_ID frames exceed the active\_connection\_id\_limit. & \textbf{S}-17 & \checkmark & Acknowledged findings.  \\  \midrule

Quinn & CVE-2023-42805: Panic when unwrapping a None value when processing an unexpected frame type. & \textbf{M}-8 & \checkmark & Fixed. \\
 & Process CRYPTO frame in 0-RTT packet. & \textbf{S}-18 & \checkmark & Fixed. \\\midrule

XQUIC~(Alibaba) & Retention of the unused encryption keys. & \textbf{S}-19 & \checkmark & Acknowledged findings. Potential security vulnerability. \textbf{Unresolved for over 90 days}. See \href{https://github.com/alibaba/xquic/issues/345}{here}.  \\ %\cline{2-4} 
 & Maintaining a number of active connection IDs that
exceed the active\_connection\_id\_limit. & \textbf{S}-20 & \checkmark & Fixed. \\ \midrule
% Reported on 7 July 2023. Fixed on 13 December 2023.

\begin{tabular}{@{}p{3cm}@{}} Aioquic, LSQUIC, Neqo, Quic-go, Quinn, Quiwi, S2n-quic~(Amazon), XQUIC \\ \\ \\ \\ \end{tabular} & \begin{tabular}{@{}p{11.5cm}@{}} Incorrect handling of the second and subsequent Initial packets carried in a UDP datagram with a payload size smaller than 1200 bytes. \\ \\ \\ \\ \\ \\ \\ \end{tabular} & \begin{tabular}{@{}c@{}} \textbf{S}-21 \\to \\ \textbf{S}-28 \\ \\ \\ \\ \\ \\ \end{tabular} &  \begin{tabular}{@{}c@{}} \checkmark \\ \\ \\ \\ \\ \\ \\ \\ \end{tabular} & \begin{tabular}{@{}p{5.4cm}@{}} Aioquic, LSQUIC, S2n-quic: Fixed. \\ However, Neqo, Quic-go and Quinn teams acknowledged the protocol violation. Fix: Developers propose amending the QUIC specification to provide futher clarifications. \\  Others: Acknowledged findings. \\ \\ \end{tabular} \\ \midrule

\begin{tabular}{@{}p{3cm}@{}} Aioquic, Kwik, MsQuic, LSQuic, Quant, Quiche, Quic-go, Quiche4j, Quiwi, S2n-quic \\ \end{tabular} & \begin{tabular}{@{}p{11.5cm}@{}} Accept Handshake packet from an unmatched Destination Connection ID. \\ \\ \\ \\ \\ \end{tabular} & \begin{tabular}{@{}c@{}} \textbf{S}-29 \\to \\ \textbf{S}-38  \\ \\ \\ \end{tabular} &  \begin{tabular}{@{}c@{}} \checkmark \\ \\ \\ \\ \\ \end{tabular} & \begin{tabular}{@{}p{5.4cm}@{}} Aioquic, Kwik, Lsquic, S2n-quic: Fixed. \\ Others: Acknowledged findings. \\ \\ \\ \\ \end{tabular} \\ \midrule

\begin{tabular}{@{}p{3cm}@{}}Lsquic, MsQuic, Neqo, Quiche4j, Quinn, XQUIC \\ \end{tabular} & \begin{tabular}{@{}p{11.5cm}@{}}Incorrect handling of packets without a frame. \\ \\ \\ \end{tabular} & \begin{tabular}{@{}c@{}} \textbf{S}-39 \\to \\ \textbf{S}-44 \end{tabular} & \begin{tabular}{@{}c@{}}\checkmark \\ \\ \\ \end{tabular} & \begin{tabular}{@{}p{5.4cm}@{}} Lsquic, Quinn, XQUIC : Fixed. \\ Others: Acknowledged findings. \\   \\ \end{tabular}\\

\bottomrule

\end{tabular}
}
\end{table*}

\begin{table*}[t!]
\centering
\caption{Symbolized QUIC messages, configuration settings and \sut{} status (connection active or closed) used by the \learner{} and the \textsf{Mapper} (here, \textit{we only mention the symbols used in the paper}). The variable \textit{$t$} represents the possible timeout and the variable \textit{$c$} represents the possible cipher suite that the \learner{} can select. The Configuration settings (options for the \mapper{} selected by the \learner{}) are within square brackets. The output symbols that show the hidden status of the \sut{} are within angle braces. In the learned models, we represent the symbols using their acronym form for brevity.}
\vspace{-3mm}
\label{tab:symbolDictionary}
\resizebox{0.95\textwidth}{!}{%
\renewcommand{\arraystretch}{1.2}
\begin{tabular}{@{}lll@{}}
\cline{1-3}
\textbf{Input Symbol} & \textbf{Acronym} & \textbf{Description} \\ \cline{1-3} 
\cline{1-3}
 \pmsg{{initialPing($t$)}}                            & \pmsg{{initPing($t$)}}  & An Initial packet with a PING frame. \\
 \pmsg{{initialConnectionClose($t$)}}                 & \pmsg{{initConClose($t$)}} & An Initial packet with a CONNECTION\_CLOSE frame.              \\
 \pmsg{{initialNoFrame($t$)}}                 & \pmsg{{initNoFr($t$)}} & An Initial packet without a frame.                     \\
 \pmsg{{initUnexpectedFrameType($t$)}}                 & \pmsg{{initUnxpFrType($t$)}} & An Initial packet with 0xFF frame type.                   \\
  \begin{tabular}{@{}l@{}}\pmsg{{initialClientHello-validACK($t$,$c$)}} \\ \\ \\ \end{tabular}            & \begin{tabular}{@{}l@{}}\pmsg{{initCltHello-vldACK($t$,$c$)}}\\ \\ \\ \end{tabular} & \begin{tabular}{@{}l@{}}An Initial packet with a CRYPTO frame carrying Client Hello message. This input \\will respond to the Server Hello message with an Initial packet with an ACK frame \\with PADDING frames.\end{tabular}       \\
 \begin{tabular}{@{}l@{}}\pmsg{{initialClientHello-invalidACK($t$,$c$)}} \\ \\ \\ \end{tabular}          & \begin{tabular}{@{}l@{}}\pmsg{{initCltHello-invldACK($t$,$c$)}} \\ \\ \\ \end{tabular} & \begin{tabular}{@{}l@{}}An Initial packet with a CRYPTO frame carrying Client Hello message. This input \\will respond to the Server Hello message with an Initial packet with an ACK frame \\with no PADDING frames.\end{tabular} \\
& & {$c \in \{AES\_128, AES\_256, ChaCha20\}$}   \\
\cline{1-3}
 \pmsg{{0rttPing($t$)}}                            & \pmsg{{0rttPing($t$)}}  & A 0-RTT packet with a PING frame. \\
 \pmsg{{0rttConnectionClose($t$)}}                 & \pmsg{{0rttConClose($t$)}} & A 0-RTT packet with a CONNECTION\_CLOSE frame.              \\
 \pmsg{{0rttNoFrame($t$)}}                 & \pmsg{{0rttNoFr($t$)}} & A 0-RTT packet without a frame.                     \\
 \pmsg{{0rttUnexpectedFrameType($t$)}}                 & \pmsg{{0rttUnxpFrType($t$)}} & A 0-RTT packet with 0xFF frame type.                   \\
 \pmsg{{0rttFinished($t$)}}                 & \pmsg{{0rttFin($t$)}} & A 0-RTT packet with a CRYPTO frame carrying Finished message.             \\
 \pmsg{{0rttACK($t$)}}                 & \pmsg{{0rttACK($t$)}} & A 0-RTT packet with an invalid ACK frame.                   \\
\cline{1-3}
 \pmsg{{handshakePing($t$)}}                          & \pmsg{{hndPing($t$)}} & A Handshake packet with a PING frame.                 \\
 \pmsg{{handshakeConnectionClose($t$)}}               & \pmsg{{hndConClose($t$)}} & A Handshake packet with a CONNECTION\_CLOSE frame.  \\
 \pmsg{{handshakeNoFrame($t$)}}                       & \pmsg{{hndNoFr($t$)}} & A Handshake packet without a frame.                    \\
 \pmsg{{handshakeUnexpectedFrameType($t$)}}           & \pmsg{{hndUnxpFrType($t$)}} & A Handshake packet with 0xFF frame type.         \\
 \pmsg{{handshakeEmptyCertificate ($t$)}}             & \pmsg{{hndEmpCert($t$)}} & A Handshake packet with a CRYPTO frame type carrying an empty list of certificates. \\
 \begin{tabular}{@{}l@{}}\pmsg{{handshakeInvalidCertificate($t$)}}  \\ \\ \end{tabular}           & \begin{tabular}{@{}l@{}}\pmsg{{hndInvldCert($t$)}} \\ \\ \end{tabular}   & \begin{tabular}{@{}l@{}}A Handshake packet with a CRYPTO frame type carrying a certificate that is not \\signed by the certificate authority used for verification.\end{tabular}  \\                    
 \begin{tabular}{@{}l@{}}\pmsg{{handshakeCertificate($t$)}} \\ \\ \end{tabular}                    & \begin{tabular}{@{}l@{}}\pmsg{{hndCert($t$)}} \\ \\ \end{tabular}  & \begin{tabular}{@{}l@{}}A Handshake packet with a CRYPTO frame type carrying a certificate that is signed \\by the certificate authority used for verification.\end{tabular}  \\
 \pmsg{{handshakeCertificateVerify($t$)}}             & \pmsg{{hndCertVer($t$)}} & A Handshake packet with a CRYPTO frame type carrying Certificate Verify message. \\
 \pmsg{{handshakeFinished($t$)}}                      & \pmsg{{hndFin($t$)}} & A Handshake packet with a CRYPTO frame type carrying Finished message. \\
\cline{1-3}
 \begin{tabular}{@{}l@{}}\pmsg{{ValidNewConnectionID($t$)}} \\ \\ \end{tabular}   & \begin{tabular}{@{}l@{}}\pmsg{{VldNewConID($t$)}} \\ \\ \end{tabular} & \begin{tabular}{@{}l@{}}A 1-RTT packet with a number of NEW\_CONNECTION\_ID frames that follows the \\number of Connection IDs that the \sut{} can support.\end{tabular} \\
 \begin{tabular}{@{}l@{}}\pmsg{{InvalidNewConnectionID($t$)}}  \\ \\ \end{tabular}  & \begin{tabular}{@{}l@{}}\pmsg{{InvldNewConID($t$)}} \\ \\ \end{tabular} & \begin{tabular}{@{}l@{}}A 1-RTT packet with a number of NEW\_CONNECTION\_ID frames that exceed the \\number of Connection IDs that the \sut{} can support.\end{tabular} \\
\cline{1-3}
 \pmsg{{{[}IncludeRetryToken{]}}}                & \pmsg{{{[}IncRetryTkn{]}}}  & Instructs the Mapper to include the Retry Token in its following Initial packets.   \\
 \pmsg{{[}RemovePaddingFromInitialPackets{]}} & \pmsg{{{[}RmPadFrmInitPkts{]}}} & Instructs the Mapper to remove PADDING frames from its following Initial packets.  \\
 \begin{tabular}{@{}l@{}}\pmsg{{[}ChangeDestinationConnectionID-Original{]}} \\ \\ \end{tabular} & \begin{tabular}{@{}l@{}}\pmsg{{{[}ChgDestConID-Ori{]}}} \\ \\ \end{tabular} & \begin{tabular}{@{}l@{}}Instructs the Mapper to change the Destination Connection ID of its following packets \\to original\_destination\_connection\_id. \end{tabular} \\ 
\bottomrule
                    \textbf{Output Symbol}                      &  \textbf{Acronym}     & \textbf{Description}            \\\cline{1-3}
 
                    \cline{1-3}
                     \pmsg{{retry}}                          & \pmsg{{retry}} &   A Retry packet that carries a Retry Token  \\

                     \cline{1-3}
                     \pmsg{{initialServerHello}}             & \pmsg{{initSvrHello}} & A Server Hello message encapsulated in a CRYPTO frame of an Initial packet.        \\
 
                     \cline{1-3}
                     \begin{tabular}{@{}l@{}}\pmsg{{handshakeEncryptedExtensions}}\\ \\ \end{tabular}   & \begin{tabular}{@{}l@{}}\pmsg{{hndEncExt}}\\ \\ \end{tabular}  & \begin{tabular}{@{}l@{}}An Encrypted Extensions message encapsulated in a CRYPTO frame of a Handshake \\packet. \end{tabular}\\
                     \begin{tabular}{@{}l@{}}\pmsg{{handshakeCertificateRequest}}\\ \\ \end{tabular}    & \begin{tabular}{@{}l@{}}\pmsg{{hndCertReq}}\\ \\ \end{tabular} & \begin{tabular}{@{}l@{}}A Certificate Request message encapsulated in a CRYPTO frame of a Handshake \\packet. \end{tabular}   \\
                     \pmsg{{handshakeCertificate}}           & \pmsg{{hndCert}} & A Certificate message encapsulated in a CRYPTO frame of a Handshake packet.      \\
                     \pmsg{{handshakeCertificateVerify}}     & \pmsg{{hndCertVer}}  & A Certificate Verify message encapsulated in a CRYPTO frame of a Handshake packet. \\
                     \pmsg{{handshakeFinished}}              & \pmsg{{hndFin}} & A Finished message encapsulated in a CRYPTO frame of a Handshake packet.        \\

                      \cline{1-3}
                     \pmsg{{HandshakeDone}}                  & \pmsg{{HndshkDone}} & A HANDSHAKE\_DONE frame in a 1-RTT packet.                        \\
                     \pmsg{{NewToken}}                      & \pmsg{{NewTkn}}  & A NEW\_TOKEN frame in a 1-RTT packet.                        \\
                     \pmsg{{ValidNewConnectionID}}           & \pmsg{{VldNewConID}}  & A NEW\_CONNECTION\_ID frame in a 1-RTT packet.  \\

                      \cline{1-3}
                     \pmsg{{PingACK}}                        & \pmsg{{PingACK}} & A PING ACK frame in either Initial, Handshake, or 1-RTT packets.                        \\
                     \pmsg{{ConnectionClose}}                & \pmsg{{ConClose}} & A CONNECTION\_CLOSE frame in either Initial, Handshake, or 1-RTT packets.                        \\

                    \cline{1-3}
                     \pmsg{{\textless{}ConnectionActive\textgreater{}}} & \pmsg{{\textless{}ConAct\textgreater{}}} & Indicates the \sut{} has closed the connection. \\
                     \pmsg{{\textless{}ConnectionClosed\textgreater{}}} & \pmsg{{\textless{}ConClosed\textgreater{}}} & Indicates the connection with the \sut{} is still active.\\\cline{1-3}
\end{tabular}%
}
\end{table*}

\end{document}